\documentclass[prd,showpacs,showkeys,floatfix,twocolumn,amsmath,amssymb,floatfix]{revtex4}
\usepackage{graphicx,color,dcolumn,booktabs,bm}
\usepackage{subfigure}
\usepackage{longtable,lscape}
\usepackage{amssymb}
\usepackage{indentfirst}
\usepackage{epsfig}
\usepackage{feynmf}   %{feynmp}
\usepackage{epstopdf}   %{feynmp}
\usepackage{slashed}  %for Feynman symbols
\usepackage{cases}
\usepackage{color}
\usepackage{multirow}
\usepackage{graphicx,color,dcolumn,booktabs,bm}
\usepackage[colorlinks, citecolor=blue,anchorcolor=red,menucolor=red, linkcolor=red,filecolor=red,runcolor=red,urlcolor=blue,frenchlinks=red]{hyperref}
\usepackage{enumitem}

\begin{document}
%=====================================================================================
%=====================================================================================
\title{A possible partner state of the $Y(2175)$}
%=====================================================================================
%=====================================================================================
%

\author{Hua-Xing Chen$^1$}
\email{hxchen@buaa.edu.cn}
\author{Cheng-Ping Shen$^1$}
\email{shencp@ihep.ac.cn}
\author{Shi-Lin Zhu$^{2,3,4}$}
\email{zhusl@pku.edu.cn}
\affiliation{
$^1$School of Physics and Beijing Key Laboratory of Advanced Nuclear Materials and Physics, Beihang University, Beijing 100191, China \\
$^2$School of Physics and State Key Laboratory of Nuclear Physics and Technology, Peking University, Beijing 100871, China \\
$^3$Collaborative Innovation Center of Quantum Matter, Beijing 100871, China \\
$^4$Center of High Energy Physics, Peking University, Beijing 100871, China}

\begin{abstract}
We study the $Y(2175)$ using the method of QCD sum rules. There are two independent $ss\bar s\bar s$ interpolating currents with $J^{PC} = 1^{--}$, and we calculate both their diagonal and off-diagonal correlation functions. We obtain two new currents which do not strongly correlate to each other, so they may couple to two different physical states: one of them couples to the $Y(2175)$, while the other may couple to another state whose mass is about $71 ^{+172}_{-~48}$ MeV larger. Evidences of the latter state can be found in the BaBar~\cite{Aubert:2007ur}, BESII~\cite{Ablikim:2007ab}, Belle~\cite{Shen:2009zze}, and BESIII~\cite{Ablikim:2014pfc} experiments.
\end{abstract}
\pacs{12.39.Mk, 11.40.Dw, 12.38.Lg, 12.40.Yx}
\keywords{exotic hadrons, interpolating currents}
\maketitle
\pagenumbering{arabic}
%
%
%
%=====================================================================================
%=====================================================================================
\section{Introduction}
\label{sec:intro}
%=====================================================================================
%=====================================================================================
%

In recent years there have been lots of exotic hadrons observed in hadron experiments~\cite{pdg}, which can not be explained in the traditional quark model and are of particular importance to understand the low energy behaviours of Quantum Chromodynamics (QCD)~\cite{Chen:2016qju,Klempt:2007cp,Lebed:2016hpi,Esposito:2016noz,Guo:2017jvc,Ali:2017jda,Olsen:2017bmm}. Most of them contain heavy quarks, such as the charmonium-like $XYZ$ states, while there are not so many exotic hadrons in the light sector only containing light $u/d/s$ quarks. The $Y(2175)$ is one of them, which is often taken as the strange analogue of the $Y(4220)$~\cite{Aubert:2005rm,Ablikim:2016qzw}.

\begin{figure*}[h]
\centering
\subfigure[The $e^+ e^- \to \phi f_{0}(980)$ cross section. Taken from BaBar~\cite{Aubert:2006bu}.]{
\label{fig:BaBar2006}
\includegraphics[width=0.4\textwidth]{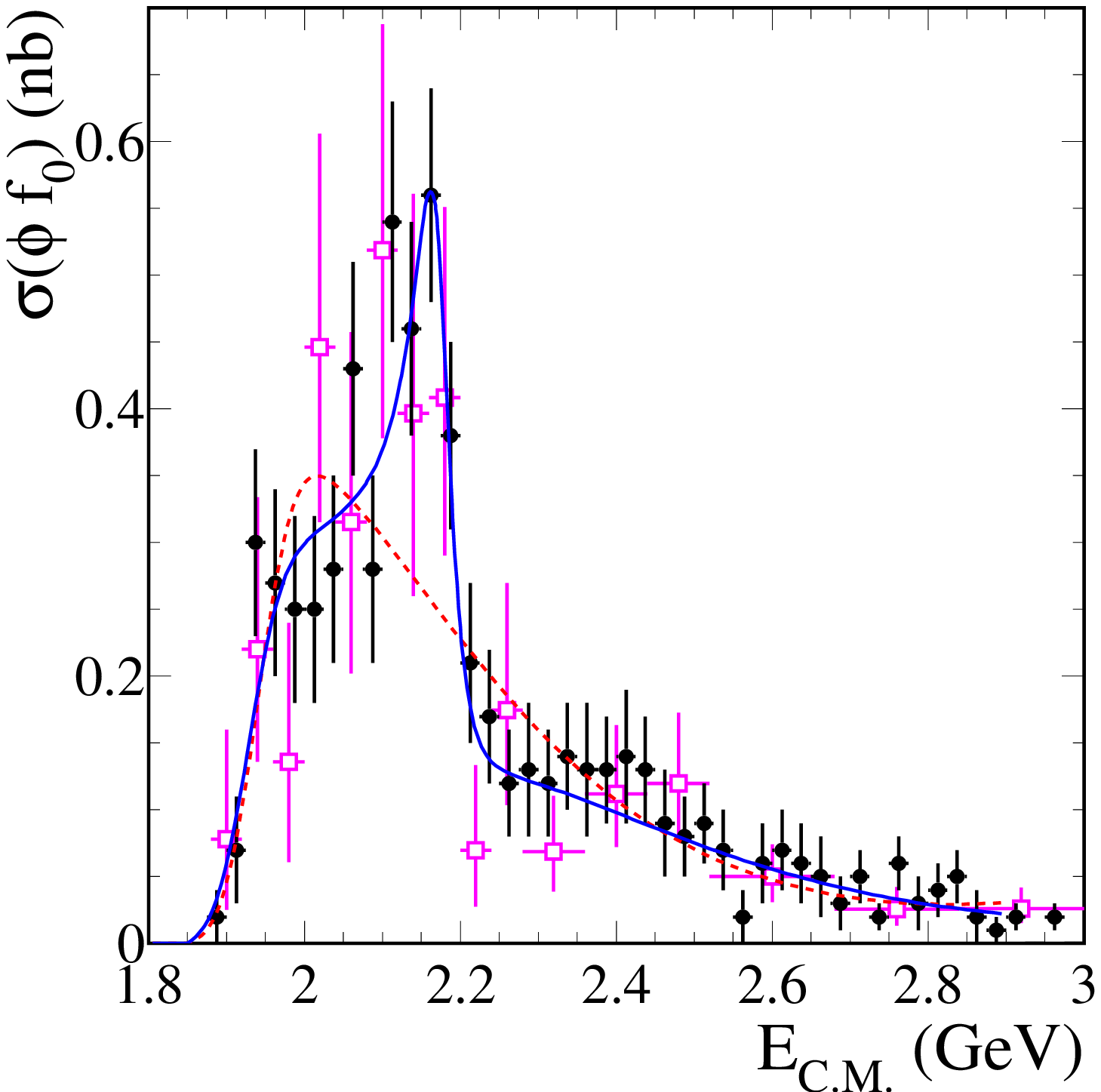}}
%%%%%%%%%%%%%%%%%%%%%%%%%%%%%%%%%%%%%%%%%
\subfigure[The $K^+ K^- \pi^+ \pi^-$ invariant mass distribution in the $K^+ K^- f_0(980)$ threshold region. The fits are done by including no (dashed), one (solid) and two (dotted) resonances. Taken from BaBar~\cite{Aubert:2007ur}.]{
\label{fig:BaBar2007}
\includegraphics[width=0.4\textwidth]{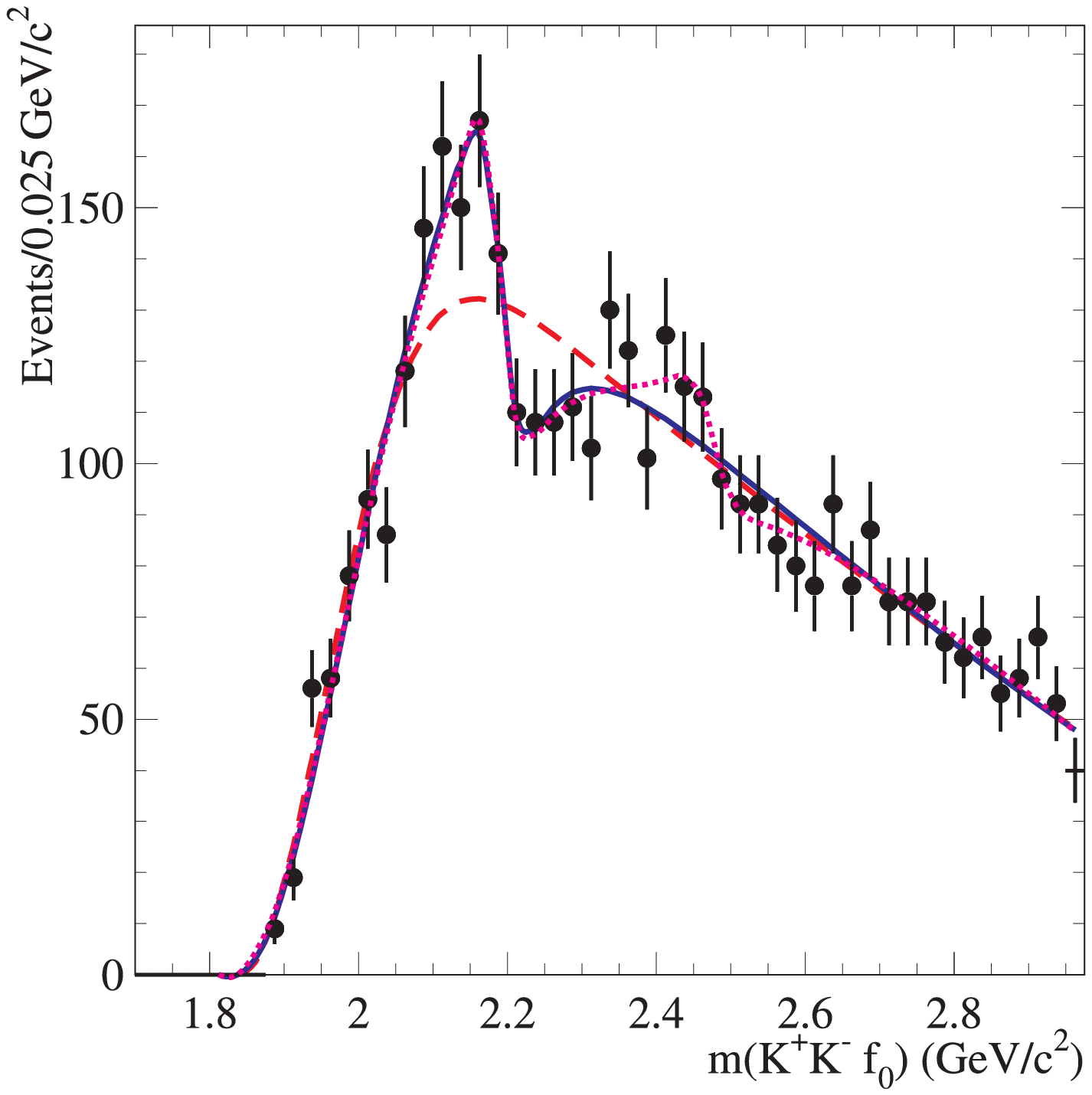}}
%%%%%%%%%%%%%%%%%%%%%%%%%%%%%%%%%%%%%%%%%
\\
\subfigure[The $e^+ e^- \to \phi \pi^+ \pi^-$ cross section with two incoherent Briet-Wigner functions, the $\phi(1680)$ and the $Y(2175)$. Taken from Belle~\cite{Shen:2009zze}.]{
\label{fig:Belle2008}
\includegraphics[width=0.4\textwidth]{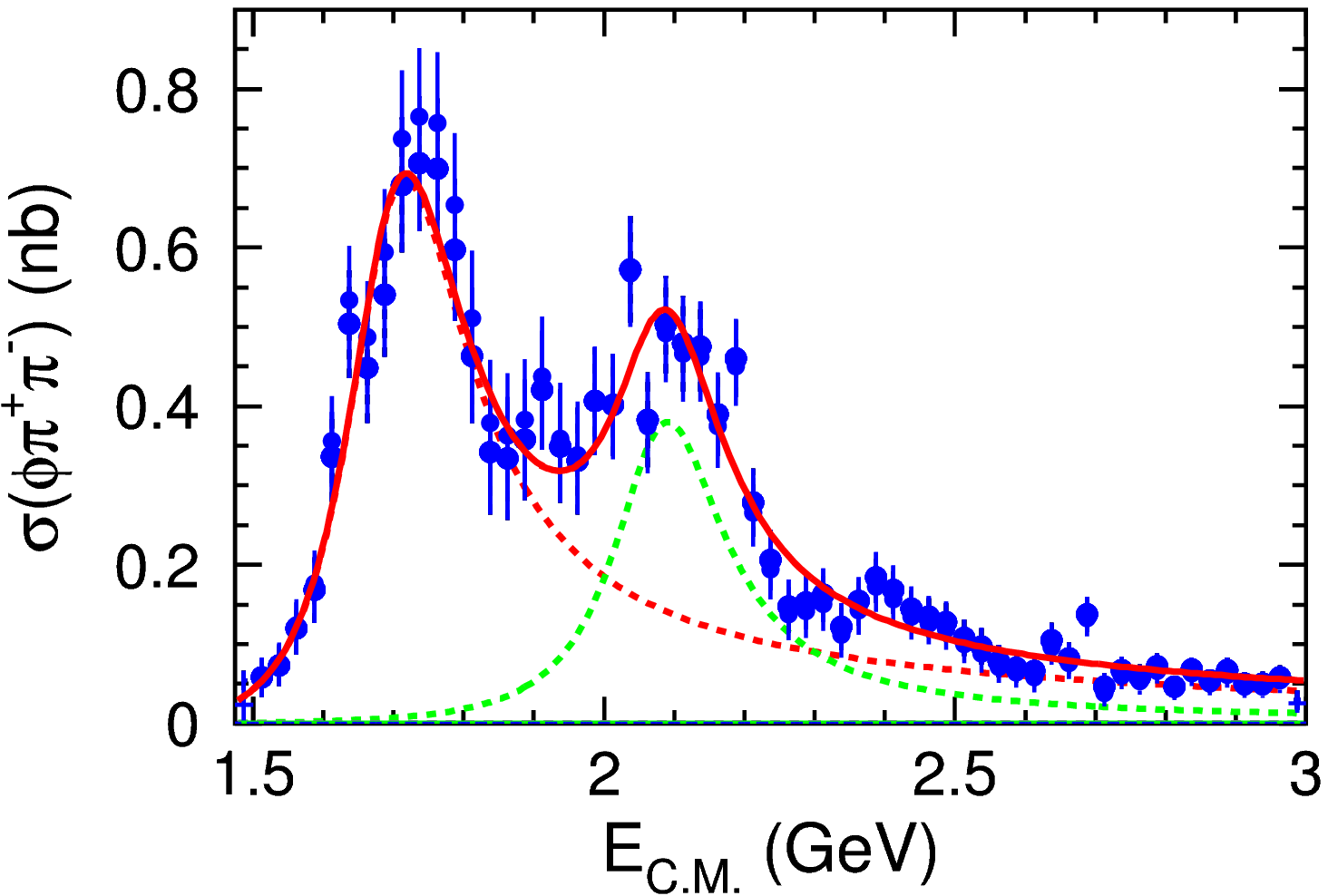}
}
%%%%%%%%%%%%%%%%%%%%%%%%%%%%%%%%%%%%%%%%%
\subfigure[Fits to the BaBar~\cite{Aubert:2006bu,Aubert:2007ur} and Belle~\cite{Shen:2009zze} measurements of the $e^+ e^- \to \phi f_0(980)$ cross sections with two coherent Briet-Wigner functions, performed by Shen and Yuan and taken from Ref.~\cite{Shen:2009mr}.]{
\label{fig:Shen2009}
\includegraphics[width=0.4\textwidth]{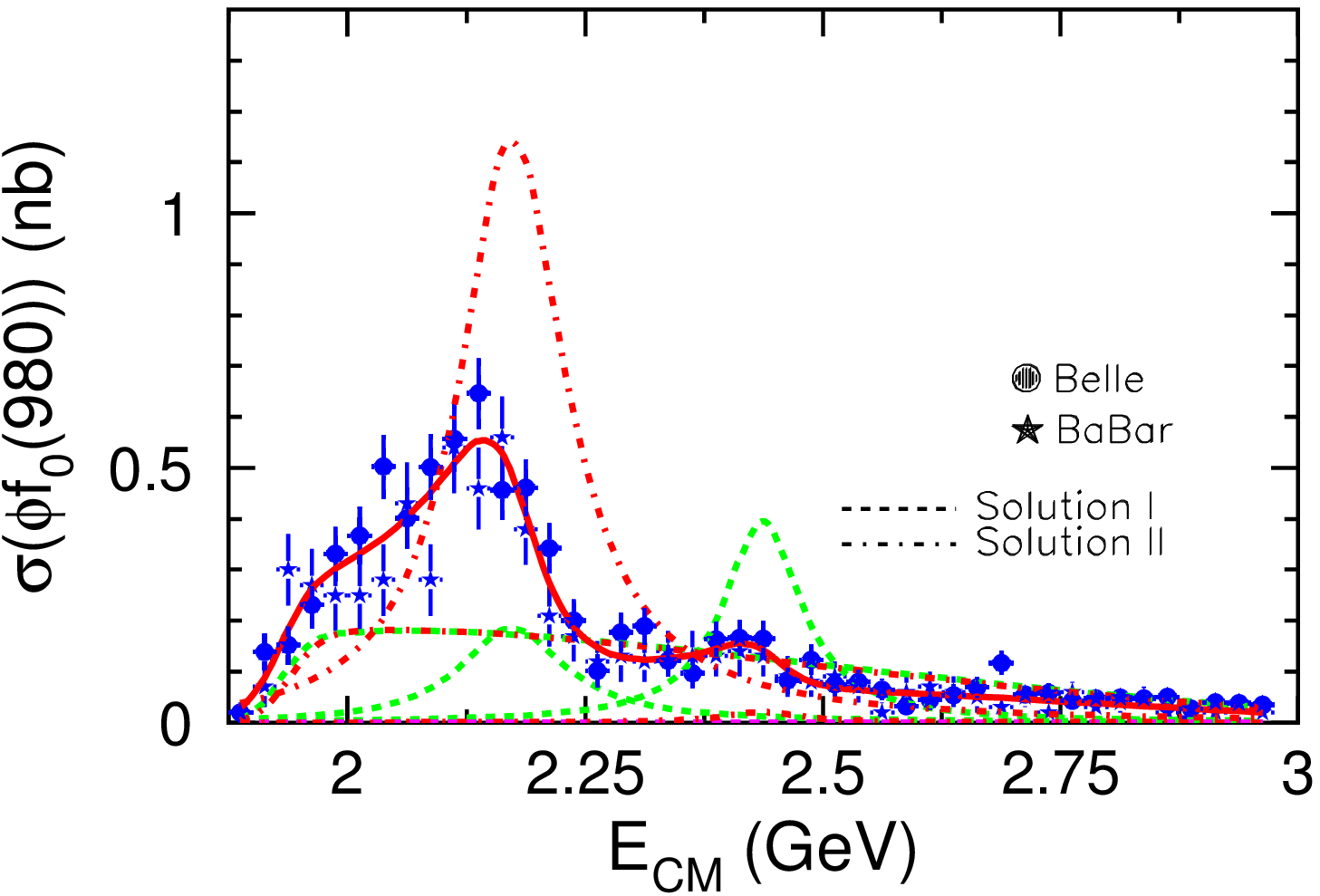}
}
%%%%%%%%%%%%%%%%%%%%%%%%%%%%%%%%%%%%%%%%%
\\
\subfigure[The $\phi f_0(980)$ invariant mass spectrum. Taken from BESII~\cite{Ablikim:2007ab}.]{
\label{fig:BES2007}
\includegraphics[width=0.4\textwidth]{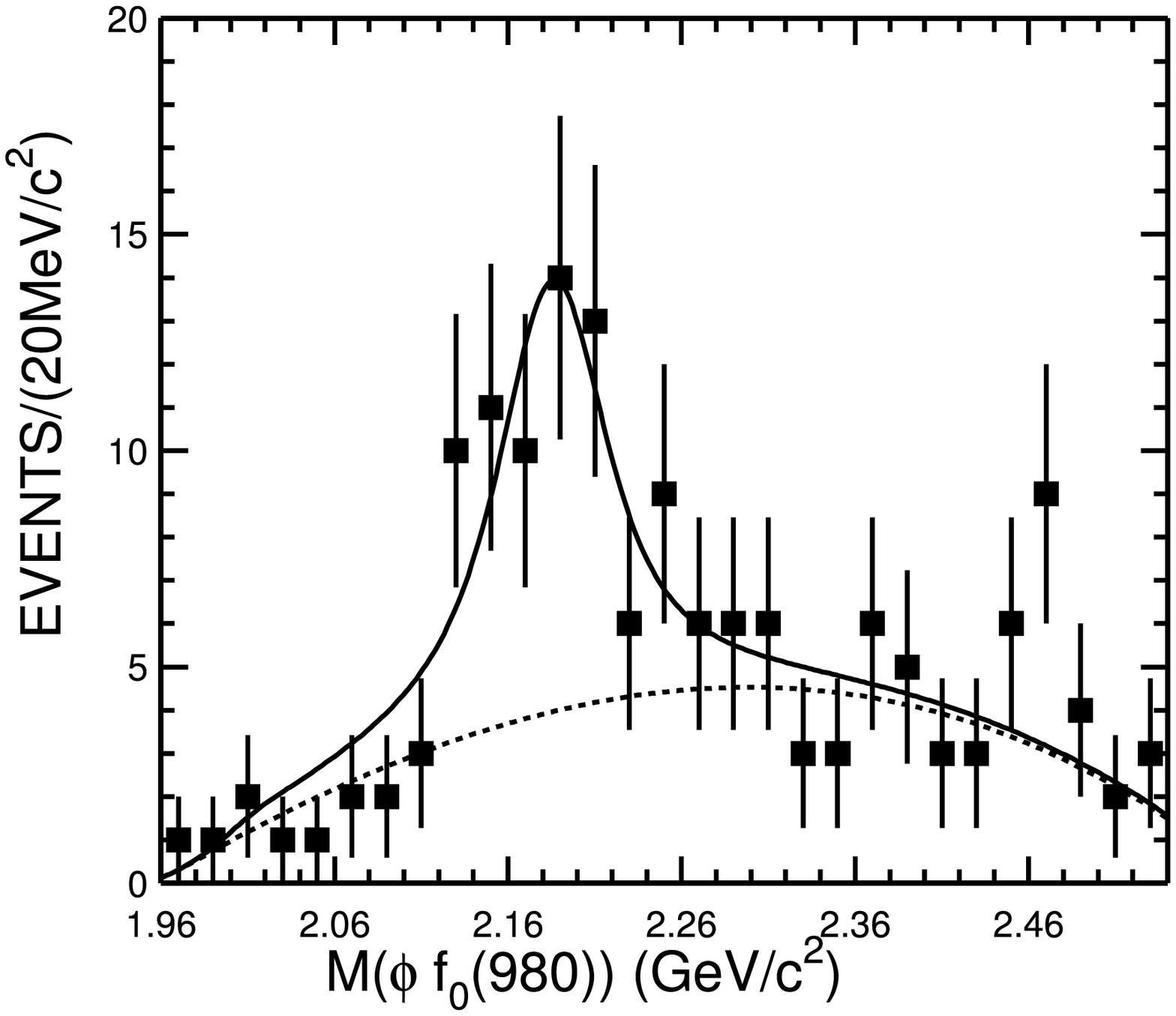}
}
%%%%%%%%%%%%%%%%%%%%%%%%%%%%%%%%%%%%%%%%%
\subfigure[The $\phi f_0(980)$ invariant mass spectrum. Taken from BESIII~\cite{Ablikim:2014pfc}.]{
\label{fig:BES2014}
\includegraphics[width=0.4\textwidth]{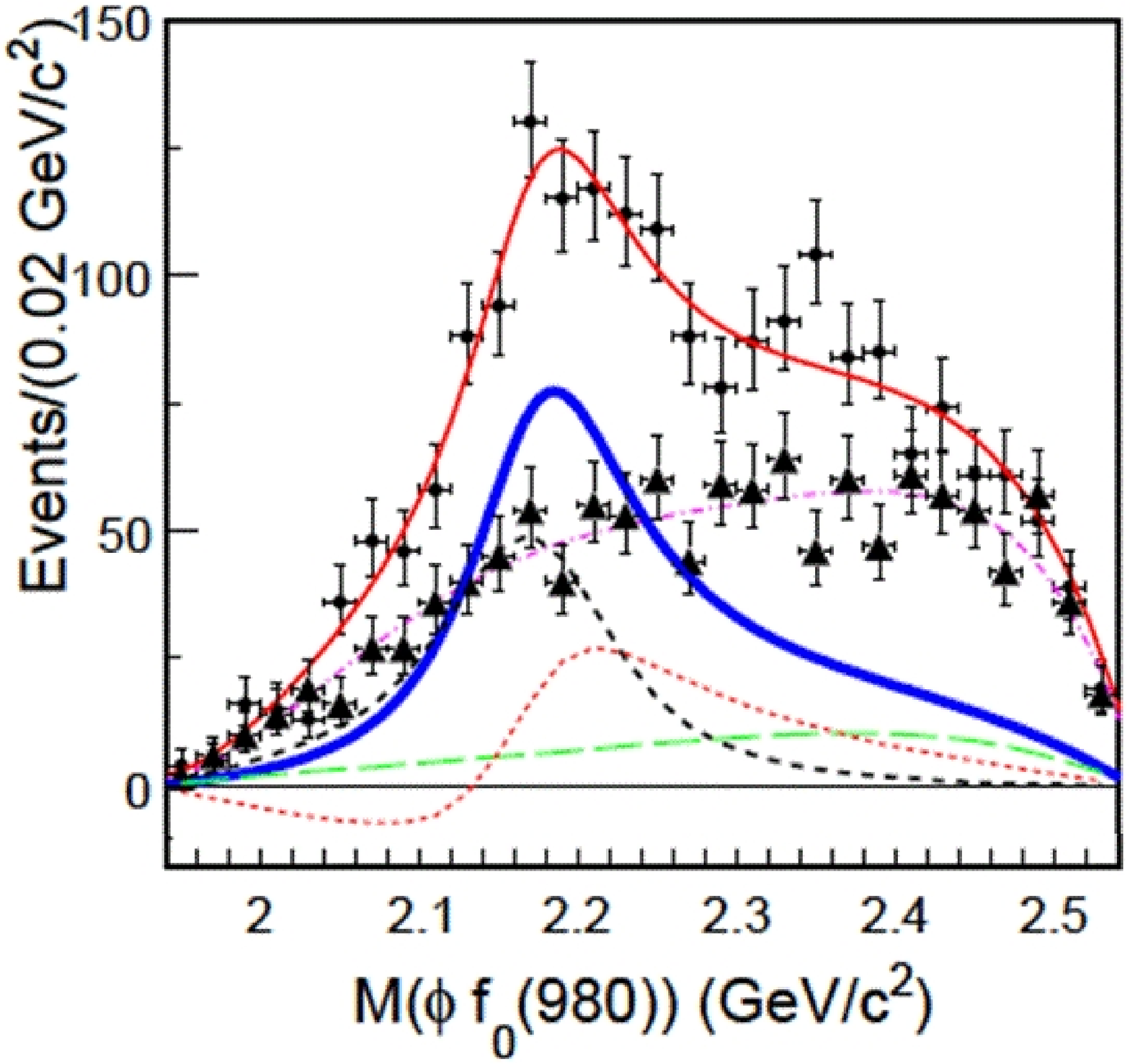}
}
\caption{
The BaBar~\cite{Aubert:2006bu,Aubert:2007ur}, Belle~\cite{Shen:2009zze}, BESII~\cite{Ablikim:2007ab}, and BESIII~\cite{Ablikim:2014pfc} experiments observing the $Y(2175)$ as well as the fit performed in Ref.~\cite{Shen:2009mr}.}
\label{fig:experiments}
\end{figure*}

The $Y(2175)$ was first observed in 2006 by the BaBar Collaboration in the $\phi f_0(980)$ invariant mass spectrum~\cite{Aubert:2006bu,Aubert:2007ur,Aubert:2007ym,Lees:2011zi}, and later confirmed in the BESII~\cite{Ablikim:2007ab}, Belle~\cite{Shen:2009zze}, and BESIII~\cite{Ablikim:2014pfc,Ablikim:2017auj} experiments. Its mass and width were measured to be $M = 2188 \pm 10$ MeV and $\Gamma = 83 \pm 12$ MeV respectively, and its spin-parity quantum number is $J^{PC} = 1^{--}$~\cite{pdg}. We list some of these experiments in Fig.~\ref{fig:experiments}, including:
\begin{enumerate}

\item[]Fig.~\ref{fig:BaBar2006}: the BaBar experiment~\cite{Aubert:2006bu} discovering the $Y(2175)$ in the $e^+ e^- \to \phi f_{0}(980)$ cross section in 2006.

\item[]Fig.~\ref{fig:BaBar2007}: the BaBar experiment~\cite{Aubert:2007ur} in 2007.

\item[]Fig.~\ref{fig:Belle2008}: the Belle experiment~\cite{Shen:2009zze} in 2009.

\item[]Fig.~\ref{fig:Shen2009}: a combined fit to the BaBar~\cite{Aubert:2006bu,Aubert:2007ur} and Belle~\cite{Shen:2009zze} measurements of the $e^+ e^- \to \phi f_{0}(980)$ cross sections, performed by Shen and Yuan in Ref.~\cite{Shen:2009mr}.

\item[]Fig.~\ref{fig:BES2007}: the BESII experiment~\cite{Ablikim:2007ab} in 2007.

\item[]Fig.~\ref{fig:BES2014}: the BESIII experiment~\cite{Ablikim:2014pfc} in 2014.
\end{enumerate}
Besides the $Y(2175)$, there might be another structure in the $\phi f_0(980)$ invariant mass spectrum at around 2.4 GeV, whose evidences can be found in the BaBar~\cite{Aubert:2007ur} (Fig.~\ref{fig:BaBar2007} around 2.4 GeV), Belle~\cite{Shen:2009zze} (Fig.~\ref{fig:Belle2008} around 2.40 GeV), BESII~\cite{Ablikim:2007ab} (Fig.~\ref{fig:BES2007} around 2.46 GeV), and BESIII~\cite{Ablikim:2014pfc} (Fig.~\ref{fig:BES2014} around 2.35 GeV) experiments. The BaBar experiment~\cite{Aubert:2007ur} determined its mass and width to be $2.47 \pm 0.07$ GeV and $77 \pm 65$ MeV, respectively. Shen and Yuan~\cite{Shen:2009mr} also used the BaBar~\cite{Aubert:2006bu,Aubert:2007ur} and Belle~\cite{Shen:2009zze} data to fit its mass and width to be $2436 \pm 34$ MeV and $99 \pm 105$ MeV, respectively. However, its statistical significance is smaller than $3.0\sigma$.
In this paper we shall study this structure as well as the $Y(2175)$ simultaneously using the method of QCD sum rules.

Since its discovery, the $Y(2175)$ has attracted much attention from the hadron physics community, and many theoretical methods and models were applied to study it. By using both the chiral unitary model~\cite{Napsuciale:2007wp,GomezAvila:2007ru} and the Faddeev equations~\cite{MartinezTorres:2008gy}, the authors interpreted the $Y(2175)$ as a dynamically generated state in the $\phi K \bar K$ and $\phi \pi \pi$ systems, and more states were predicted in the $\phi \pi^0 \eta$~\cite{VaqueraAraujo:2009jq} and $N K \bar K$~\cite{MartinezTorres:2009cw,Xie:2010ig} systems. By using similar approaches, the $Y(2175)$ was interpreted as a dynamically generated resonance by the self-interactions between the $\phi$ and $f_0(980)$ resonances~\cite{AlvarezRuso:2009xn}, while the resonance spectrum expansion formalism by including the $f_0(980)$ as a resonance in the coupled $\pi \pi$-$KK$ system is also able to generate the $Y(2175)$ in the $\phi f_0(980)$ channel~\cite{Coito:2009na}.

Besides the dynamically generated resonance, there are many other interpretations to explain this structure. In Ref.~\cite{Ding:2007pc} the authors interpreted the $Y(2175)$ as a $2^3D_1$ $s\bar{s}$ meson, and calculated its decay modes using both the $^{3}P_0$ model and the flux-tube model. In Ref.~\cite{Abud:2009rk} the authors used a constituent quark model to interpret the $Y(2175)$ as a hidden-strangeness baryon-antibaryon state ($q q s \bar q \bar q \bar s$) strongly coupling to the $\Lambda \bar \Lambda$ channel. Later in Ref.~\cite{Zhao:2013ffn} the authors applied the one-boson-exchange model to interpret the $Y(2175)$ and $\eta(2225)$ as the bound states of $\Lambda \bar \Lambda(^3S_1)$ and $\Lambda \bar \Lambda(^1S_0)$, respectively. In Ref.~\cite{Ding:2006ya} the authors interpreted the $Y(2175)$ as a strangeonium hybrid state and used the flux-tube model to study its decay properties. However, this interpretation is not supported by the non-perturbative lattice QCD calculations~\cite{Dudek:2011bn}. Productions of the $Y(2175)$ were studied in Refs.~\cite{Bystritskiy:2007wq,Ali:2011qi} by using the Nambu-Jona-Lasinio model and the Drell-Yan mechanism, while its decay properties were studied in Refs.~\cite{Chen:2011cj,Wang:2012wa} via the initial single pion emission mechanism.

The method of QCD sum rules was also applied to study the $Y(2175)$~\cite{Wang:2006ri,Chen:2008ej}, which method has been widely and successfully used to study hadron properties~\cite{Shifman:1978bx,Reinders:1984sr}. When using this method to investigate a physical state, one needs to construct the relevant interpolating current, but we still do not fully understand their relations: a) the interpolating current sees only the quantum numbers of the physical state, so it can also couple to some other physical states as well as the relevant threshold; b) one can sometimes construct more than one interpolating currents, all of which can couple to the same physical state. Some previous studies tell us that:
\begin{enumerate}

\item In Ref.~\cite{Chen:2015kpa} we systematically studied the $P$-wave singly heavy baryons. Theoretically, we find that they have rich internal structures, %, and for each structure we can construct one interpolating current.
and there can be as many as three excited $\Omega_c$ states of $J^P =1/2^-$, three of $J^P =3/2^-$, and one of $J^P =5/2^-$. For each state we can construct one interpolating current having the same internal structure. The numbers of excited $\Lambda_c/\Xi_c/\Sigma_c/\Xi_c^\prime$ states/currents are the same.

    We do not know which of them exist in nature, but we do know that, experimentally, the $P$-wave charmed baryons also have rich structures~\cite{Chen:2016spr}, for example, the LHCb experiment~\cite{Aaij:2017nav} observed as many as five excited $\Omega_c$ states at the same time, all of which can be $P$-wave charmed baryons.

%\item In Ref.~\cite{Chen:2017dpy} the mass spectra of hidden-charm tetraquark states with $J^{PC} = 0^{++}$ and $2^{++}$ were systematically investigated. There can be as many as ten $q c \bar q \bar c$ ($q=u/d$) interpolating currents of %$J^{PC} = 0^{++}$, and four of $J^{PC} = 2^{++}$.
%
%    Again, we do not know which of them exist, but it is unbelievable if all of them exist in nature. Their possible candidates observed in experiments so far are the $X(3860)$, $X(3915)$, $X(4160)$, and $X(4350)$, but there are many other %possible assignments for these states.

\item In Ref.~\cite{Chen:2010ze} the mass spectra of vector and axial-vector hidden-charm tetraquark states were systematically investigated. There can be as many as eight $q c \bar q \bar c$ ($q=u/d$) interpolating currents of $J^{PC} = 1^{--}$. Comparably, there have been many vector charmonium-like states observed in hadron experiments~\cite{pdg}, including the $Y(4008)$~\cite{Yuan:2007sj}, $Y(4220)$~\cite{Aubert:2005rm}, $Y(4320)$~\cite{Ablikim:2016qzw}, $Y(4360)$~\cite{Aubert:2007zz}, $Y(4630)$~\cite{Pakhlova:2008vn}, and $Y(4660)$~\cite{Wang:2007ea}, etc.

\item In Ref.~\cite{Chen:2016otp} we systematically studied hidden-charm pentaquark states having spin $J={1\over2}/{3\over2}/{5\over2}$. We constructed hundreds of hidden-charm pentaquark interpolating currents, suggesting that their internal structures are rather complicated. However, the only observed hidden-charm pentaquark states so far are the $P_c(4380)$ and $P_c(4450)$~\cite{Aaij:2015tga}, and it is unbelievable that there exist hundreds of hidden-charm pentaquark states in nature.

\end{enumerate}
Hence, the internal structures of (exotic) hadrons are complicated. For each internal structure we can construct the relevant interpolating current, and their relations are also complicated. Especially, there can be many interpolating currents when studying exotic hadrons, which makes them not easy to handle.

To clarify this problem, a good subject is to study the $Y(2175)$ of $J^{PC} = 1^{--}$. The relevant $ss\bar s \bar s$ interpolating currents have been systematically constructed in Ref.~\cite{Chen:2008ej}, and there are only two independent ones. We have separately used them to perform QCD sum rule analyses, both of which can be used to explain the $Y(2175)$. However, in Ref.~\cite{Chen:2008ej} we only calculated the diagonal terms of these two currents, and in this work we shall further calculate their off-diagonal term to study their correlation. This can significantly improve our understanding on the relations between interpolating currents and physical states.

Another advantage to study the $Y(2175)$ is that, experimentally, there might be another structure in the $\phi f_0(980)$ invariant mass spectrum at around 2.4 GeV, as we have discussed before. It is quite interesting to study the relations between the two independent $ss\bar s \bar s$ currents with $J^{PC} = 1^{--}$ and the two possible structures in the $\phi f_0(980)$ invariant mass spectrum, both theoretically and experimentally, and both coherently and incoherently. Note that there are many charmonium-like $Y$ states of $J^{PC} = 1^{--}$, so it is natural to think that there can be more than one $Y$ states in the light sector.

This paper is organized as follows.
In Sec.~\ref{sec:current}, we list the two independent $ss\bar s \bar s$ interpolating currents with $J^{PC} = 1^{--}$, and discuss how to diagonalize them.
In Sec.~\ref{sec:sumrule}, we use two diquark-antidiquark $(ss) (\bar s \bar s)$ interpolating currents to perform QCD sum rule analyses, and obtain two new currents which do not strongly correlate to each other.
In Sec.~\ref{sec:numerical}, we use these two new currents to calculate mass spectra, and Sec.~\ref{sec:summary} is a summary.

%
%=====================================================================================
%=====================================================================================
\section{Interpolating currents and their relations to possible physical states}
\label{sec:current}
%=====================================================================================
%=====================================================================================
%

The interpolating currents having the quark content $ss\bar s \bar s$ and with the quantum number $J^{PC} = 1^{--}$ have been systematically constructed in Ref.~\cite{Chen:2008ej}. We briefly summarize the results here and discuss their relations to possible physical states.
\begin{enumerate}

\item There are two non-vanishing diquark-antidiquark $(ss) (\bar s \bar s)$ interpolating currents with $J^{PC} = 1^{--}$:
%
%%%%%%%%%%%%%%%%%%%%%%%%%%%%%%%%%%%%%%%%%%%%%%%%%%%%%%%%%%%%%%%%%%%%%%%%%%%%%%
\begin{eqnarray}
%-------------------------------------eta 1------------------------------------
&& \eta_{1\mu} =
\\ \nonumber && ~~~ (s_a^T C \gamma_5 s_b) (\bar{s}_a \gamma_\mu
\gamma_5 C \bar{s}_b^T) - (s_a^T C \gamma_\mu \gamma_5 s_b)
(\bar{s}_a \gamma_5 C \bar{s}_b^T) \label{def:eta1} \, ,
%-------------------------------------eta 2------------------------------------
\\ && \eta_{2\mu} =
\\ \nonumber && ~~~ (s_a^T C \gamma^\nu s_b) (\bar{s}_a \sigma_{\mu\nu} C \bar{s}_b^T)
- (s_a^T C \sigma_{\mu\nu} s_b) (\bar{s}_a \gamma^\nu C \bar{s}_b^T)
\label{def:eta2} \, ,
\end{eqnarray}
%%%%%%%%%%%%%%%%%%%%%%%%%%%%%%%%%%%%%%%%%%%%%%%%%%%%%%%%%%%%%%%%%%%%%%%%%%%%%%
%
where $a$ and $b$ are color indices, $C = i\gamma_2 \gamma_0$ is the charge conjugation operator, and the superscript $T$ represents the transpose of Dirac indices.
These two currents are independent of each other.

\item There are four non-vanishing meson-meson $(\bar ss) (\bar s s)$ interpolating currents with $J^{PC} = 1^{--}$:
%
%%%%%%%%%%%%%%%%%%%%%%%%%%%%%%%%%%%%%%%%%%%%%%%%%%%%%%%%%%%%%%%%%%%%%%%%%%%%%%
\begin{eqnarray}
%-------------------------------------eta 3------------------------------------
\eta_{3\mu} &=& (\bar{s}_a s_a)(\bar{s}_b \gamma_\mu s_b) \, ,
%-------------------------------------eta 4------------------------------------
\\ \eta_{4\mu} &=& (\bar{s}_a \gamma^\nu\gamma_5 s_a)(\bar{s}_b \sigma_{\mu\nu}\gamma_5 s_b) \, ,
%-------------------------------------eta 5------------------------------------
\\ \eta_{5\mu} &=& {\lambda_{ab}}{\lambda_{cd}}(\bar{s}_a s_b)(\bar{s}_c \gamma_\mu s_d) \, ,
%-------------------------------------eta 6------------------------------------
\\ \eta_{6\mu} &=& {\lambda_{ab}}{\lambda_{cd}} (\bar{s}_a \gamma^\nu\gamma_5 s_b)(\bar{s}_c \sigma_{\mu\nu}\gamma_5 s_d) \, .
\end{eqnarray}
%%%%%%%%%%%%%%%%%%%%%%%%%%%%%%%%%%%%%%%%%%%%%%%%%%%%%%%%%%%%%%%%%%%%%%%%%%%%%%
%
However, only two of them are independent.

\item When using local currents, we can verify the following relations between the above $(ss)(\bar s\bar s)$ and $(\bar s s)(\bar s s)$ currents through the Fierz transformation:
%
%%%%%%%%%%%%%%%%%%%%%%%%%%%%%%%%%%%%%%%%%%%%%%%%%%%%%%%%%%%%%%%%%%%%%%%%%%%%%%
\begin{eqnarray}
\eta_{1\mu}(x) &=& - \eta_{3\mu}(x) + i \eta_{4\mu}(x) \, ,
\label{eq:transform}
\\ \nonumber \eta_{2\mu}(x) &=& 3i \eta_{3\mu}(x) - \eta_{4\mu}(x) \, .
\end{eqnarray}
%%%%%%%%%%%%%%%%%%%%%%%%%%%%%%%%%%%%%%%%%%%%%%%%%%%%%%%%%%%%%%%%%%%%%%%%%%%%%%
%

\end{enumerate}

In Ref.~\cite{Chen:2008ej} we have separately used $\eta_{1\mu}$ and $\eta_{2\mu}$ to perform QCD sum rule analyses, i.e., we have calculated the diagonal terms:
\begin{eqnarray}
\langle 0 | T \eta_{1\mu}(x) { \eta_{1\nu}^\dagger } (0) | 0 \rangle~~~{\rm and }~~~\langle 0 | T \eta_{2\mu}(x) { \eta_{2\nu}^\dagger } (0) | 0 \rangle \, .
\end{eqnarray}
However, although $\eta_{1\mu}$ and $\eta_{2\mu}$ are independent of each other, they can be correlated to each other, i.e., the off-diagonal term can be non-zero:
\begin{eqnarray}
\langle 0 | T \eta_{1\mu}(x) { \eta_{2\nu}^\dagger } (0) | 0 \rangle \neq 0 \, ,
\end{eqnarray}
suggesting that $\eta_{1\mu}$ and $\eta_{2\mu}$ may couple to the same physical state.
In this paper we shall evaluate this off-diagonal term in order to find two non-correlated currents:
\begin{eqnarray}
J_{1\mu} &=& \cos\theta~\eta_{1\mu} + \sin\theta~i~\eta_{2\mu} \, ,
\\ \nonumber J_{2\mu} &=& \sin\theta~\eta_{1\mu} + \cos\theta~i~\eta_{2\mu} \, ,
\end{eqnarray}
satisfying
\begin{eqnarray}
&& \langle 0 | T J_{1\mu}(x) { J_{2\nu}^\dagger } (0) | 0 \rangle = 0 \, ,
\label{eq:offdiagonal}
\\ \nonumber && ~~~~~~~~~~~~~~~~ {\rm or} ~~~
\left\{
\begin{array}{c}
\ll \langle 0 | T J_{1\mu}(x) { J_{1\nu}^\dagger } (0) | 0 \rangle
\\ \ll \langle 0 | T J_{2\mu}(x) { J_{2\nu}^\dagger } (0) | 0 \rangle
\end{array}
\right.\, .
\end{eqnarray}
Then we shall use $J_{1\mu}$ and $J_{2\mu}$ to perform QCD sum rule analyses. Due to the above Eq.~(\ref{eq:offdiagonal}), $J_{1\mu}$ and $J_{2\mu}$ should not strongly couple to the same physical state, so we assume
\begin{eqnarray}
\langle 0| J_{1\mu} | Y_1 \rangle &=& f_1~\epsilon_{\mu} \, ,
\\ \langle 0| J_{2\mu} | Y_2 \rangle &=& f_2~\epsilon_{\mu} \, ,
\end{eqnarray}
where $Y_1$ and $Y_2$ are two different states with $J^{PC} = 1^{--}$, $f_1$ and $f_2$ are decay constants, and $\epsilon_{\mu}$ is the polarization vector.
Especially, we shall evaluate the mass splitting between these two states/currents. %, through which we can improve our understanding on the relations between interpolating currents and physical states.

%
%=====================================================================================
%=====================================================================================
\section{QCD sum rule Analysis}
\label{sec:sumrule}
%=====================================================================================
%=====================================================================================
%

The method of QCD sum rules is a powerful and successful non-perturbative method~\cite{Shifman:1978bx,Reinders:1984sr}.
In this method, we calculate the two-point correlation function
%
%%%%%%%%%%%%%%%%%%%%%%%%%%%%%%%%%%%%%%%%%%%%%%%%%%%%%%%%%%%%%%%%%%%%%%%%%%%%%%
\begin{equation}
\Pi_{\mu\nu}(q^2) \, \equiv \, i \int d^4x e^{iqx} \langle 0 | T \eta_\mu(x) { \eta_\nu^\dagger } (0) | 0 \rangle \, ,
\label{def:pi}
\end{equation}
%%%%%%%%%%%%%%%%%%%%%%%%%%%%%%%%%%%%%%%%%%%%%%%%%%%%%%%%%%%%%%%%%%%%%%%%%%%%%%
%
at both the hadron and quark-gluon levels.

At the hadron level we simplify its Lorentz structure to be:
%
%%%%%%%%%%%%%%%%%%%%%%%%%%%%%%%%%%%%%%%%%%%%%%%%%%%%%%%%%%%%%%%%%%%%%%%%%%%%%%
\begin{equation}
\Pi_{\mu\nu}(q^2) = ( {q_\mu q_\nu \over q^2} - g_{\mu\nu} ) \Pi(q^2) + {q_\mu q_\nu \over q^2} \Pi^{(0)}(q^2) \, ,
\label{def:pi1}
\end{equation}
%%%%%%%%%%%%%%%%%%%%%%%%%%%%%%%%%%%%%%%%%%%%%%%%%%%%%%%%%%%%%%%%%%%%%%%%%%%%%%
%
and express $\Pi(q^2)$ in the form of the dispersion relation:
%
%%%%%%%%%%%%%%%%%%%%%%%%%%%%%%%%%%%%%%%%%%%%%%%%%%%%%%%%%%%%%%%%%%%%%%%%%%%%%%
\begin{equation}
\Pi(q^2)=\int^\infty_{16 m_s^2}\frac{\rho(s)}{s-q^2-i\varepsilon}ds \, .
\label{eq:disper}
\end{equation}
%%%%%%%%%%%%%%%%%%%%%%%%%%%%%%%%%%%%%%%%%%%%%%%%%%%%%%%%%%%%%%%%%%%%%%%%%%%%%%
%
Here $\rho(s)$ is the spectral density, for which we adopt a parametrization of one
pole dominance for the ground state $Y$ and a continuum contribution:
%
%%%%%%%%%%%%%%%%%%%%%%%%%%%%%%%%%%%%%%%%%%%%%%%%%%%%%%%%%%%%%%%%%%%%%%%%%%%%%%
\begin{eqnarray}
\rho(s) & \equiv & \sum_n\delta(s-M^2_n)\langle 0|\eta|n\rangle\langle n|{\eta^\dagger}|0\rangle
\ \nonumber\\ &=& f^2_Y\delta(s-M^2_Y)+ \rm{continuum}\, .
\label{eq:rho}
\end{eqnarray}
%%%%%%%%%%%%%%%%%%%%%%%%%%%%%%%%%%%%%%%%%%%%%%%%%%%%%%%%%%%%%%%%%%%%%%%%%%%%%%
%

At the quark-gluon level, we insert $J_{1\mu}$ and $J_{2\mu}$ into Eq.~(\ref{def:pi}), and calculate the correlation function using the method of operator product expansion (OPE).
After performing the Borel transformation at both the hadron and quark-gluon levels, we obtain
%
%%%%%%%%%%%%%%%%%%%%%%%%%%%%%%%%%%%%%%%%%%%%%%%%%%%%%%%%%%%%%%%%%%%%%%%%%%%%%%
\begin{equation}
\Pi^{(all)}(M_B^2)\equiv\mathcal{B}_{M_B^2}\Pi(p^2)=\int^\infty_{16 m_s^2} e^{-s/M_B^2} \rho(s)ds \, .
\label{eq_borel}
\end{equation}
%%%%%%%%%%%%%%%%%%%%%%%%%%%%%%%%%%%%%%%%%%%%%%%%%%%%%%%%%%%%%%%%%%%%%%%%%%%%%%
%
After approximating the continuum using the spectral density of OPE above a threshold value $s_0$, we obtain the sum rule equation
%
%%%%%%%%%%%%%%%%%%%%%%%%%%%%%%%%%%%%%%%%%%%%%%%%%%%%%%%%%%%%%%%%%%%%%%%%%%%%%%
\begin{equation}
\Pi(M_B^2) \equiv f^2_Y e^{-M_Y^2/M_B^2} = \int^{s_0}_{16 m_s^2} e^{-s/M_B^2}\rho(s)ds
\label{eq_fin} \, .
\end{equation}
%%%%%%%%%%%%%%%%%%%%%%%%%%%%%%%%%%%%%%%%%%%%%%%%%%%%%%%%%%%%%%%%%%%%%%%%%%%%%%
%
We can use this equation to calculate $M_Y$ through
%
%%%%%%%%%%%%%%%%%%%%%%%%%%%%%%%%%%%%%%%%%%%%%%%%%%%%%%%%%%%%%%%%%%%%%%%%%%%%%%
\begin{equation}
M^2_Y = \frac{\frac{\partial}{\partial(-1/M_B^2)}\Pi(M_B^2)}{\Pi(M_B^2)}
= \frac{\int^{s_0}_{16 m_s^2} e^{-s/M_B^2}s\rho(s)ds}{\int^{s_0}_{16 m_s^2} e^{-s/M_B^2}\rho(s)ds} \, .
\label{eq_LSR}
\end{equation}
%%%%%%%%%%%%%%%%%%%%%%%%%%%%%%%%%%%%%%%%%%%%%%%%%%%%%%%%%%%%%%%%%%%%%%%%%%%%%%
%

\begin{figure}[hbt]
\begin{center}
\includegraphics[width=0.3\textwidth]{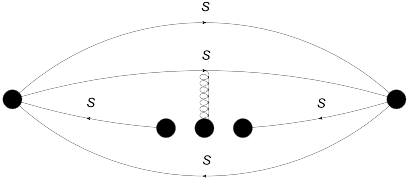}
\caption{Feynman diagram related to the quark-gluon mixed condensate $\langle g_s \bar s \sigma G s \rangle$.}
\label{fig:QG}
\end{center}
\end{figure}

The sum rules for the currents $\eta_{1\mu}$ and $\eta_{2\mu}$ have been separately calculated and given in Eqs. (13) and (14) of Ref.~\cite{Chen:2008ej}. In this paper we revise these calculations by adding the diagram shown in Fig.~\ref{fig:QG}. We write them as $\Pi_{\eta_1\eta_1}(q^2)$ and $\Pi_{\eta_2\eta_2}(q^2)$ in the present study, which are transformed to be $\Pi_{\eta_1\eta_1}(M_B^2)$ and $\Pi_{\eta_2\eta_2}(M_B^2)$ after the Borel transformation. %, and have been given in Eqs. (13) and (14) of Ref.~\cite{Chen:2008ej}.
The results are shown in Eqs.~(\ref{eq:pieta1}) and (\ref{eq:pieta2}), which do not change significantly compared to Ref.~\cite{Chen:2008ej}.

%
%%%%%%%%%%%%%%%%%%%%%%%%%%%%%%%%%%%%%%%%%%%%%%%%%%%%%%%%%%%%%%%%%%%%%%%%%%%%%%
%---------figure current 1
%%%%%%%%%%%%%%%%%%%%%%%%%%%%%%%%%%%%%%%%%%%%%%%%%%%%%%%%%%%%%%%%%%%%%%%%%%%%%%
%

In Eqs.~(\ref{eq:pieta1}) and (\ref{eq:pieta2}) we have calculated the OPE up to twelve dimension, including the strange quark mass, the perturbative term, the quark condensate $\langle \bar s s \rangle$, the gluon condensate $\langle g_s^2 GG \rangle$, the quark-gluon mixed condensate $\langle g_s \bar s \sigma G s \rangle$, and their combinations $\langle \bar s s \rangle^2$, $\langle \bar s s \rangle^3$, $\langle \bar s s \rangle^4$, $\langle g_s \bar s \sigma G s \rangle^2$, $\langle \bar s s \rangle \langle g_s \bar s \sigma G s \rangle$, $\langle \bar s s \rangle^2 \langle g_s \bar s \sigma G s \rangle$, $\langle g_s^2 GG \rangle \langle \bar s s \rangle$, $\langle g_s^2 GG \rangle \langle \bar s s \rangle^2$, $\langle g_s^2 GG \rangle \langle g_s \bar s \sigma G s \rangle$, and $\langle g_s^2 GG \rangle \langle \bar s s \rangle \langle g_s \bar s \sigma G s \rangle$. These parameters take the following values~\cite{Yang:1993bp,Narison:2002pw,Gimenez:2005nt,Jamin:2002ev,Ioffe:2002be,Ovchinnikov:1988gk,Ellis:1996xc,pdg}:
%
%%%%%%%%%%%%%%%%%%%%%%%%%%%%%%%%%%%%%%%%%%%%%%%%%%%%%%%%%%%%%%%%%%%%%%%%%%%%%%
\begin{eqnarray}
\nonumber &&\langle\bar qq \rangle=-(0.24 \pm 0.01 \mbox{ GeV})^3\, ,
\\
\nonumber &&\langle\bar ss\rangle=-(0.8\pm 0.1)\times(0.240 \mbox{
GeV})^3\, ,
\\
\nonumber &&\langle g_s^2GG\rangle =(0.48\pm 0.14) \mbox{ GeV}^4\, ,
\\ \label{condensates} && \langle g_s\bar q\sigma G q\rangle=-M_0^2\times\langle\bar qq\rangle\, ,
\\
\nonumber && M_0^2=(0.8\pm0.2)\mbox{ GeV}^2\, ,
\\
\nonumber &&m_s(2\mbox{ GeV})= 96 ^{+8}_{-4} \mbox{ MeV}\, ,
\\
\nonumber && \alpha_s(1.7\mbox{GeV}) = 0.328 \pm 0.03
\pm 0.025 \, .
\end{eqnarray}
%%%%%%%%%%%%%%%%%%%%%%%%%%%%%%%%%%%%%%%%%%%%%%%%%%%%%%%%%%%%%%%%%%%%%%%%%%%%%%
%

Beside the diagonal terms $\Pi_{\eta_1\eta_1}(q^2)$ and $\Pi_{\eta_2\eta_2}(q^2)$, in the present study we also calculate the sum rules for the off-diagonal term:
%
%%%%%%%%%%%%%%%%%%%%%%%%%%%%%%%%%%%%%%%%%%%%%%%%%%%%%%%%%%%%%%%%%%%%%%%%%%%%%%
\begin{eqnarray}
\Pi_{\mu\nu}^{\eta_1\eta_2}(q^2) &=& i \int d^4x e^{iqx} \langle 0 | T \eta_{1\mu}(x) { \eta_{2\nu}^\dagger } (0) | 0 \rangle
\\ \nonumber &=& ( {q_\mu q_\nu \over q^2} - g_{\mu\nu} ) \Pi_{\eta_1\eta_2}(q^2) + {q_\mu q_\nu \over q^2} \Pi^{(0)}_{\eta_1\eta_2}(q^2) \, .
\end{eqnarray}
%%%%%%%%%%%%%%%%%%%%%%%%%%%%%%%%%%%%%%%%%%%%%%%%%%%%%%%%%%%%%%%%%%%%%%%%%%%%%%
%
After performing the Borel transformation to $\Pi_{\eta_1\eta_2}(q^2)$, we obtain $\Pi_{\eta_1\eta_2}(M_B^2)$ as shown in Eq.~(\ref{eq:eta12}).
%
%%%%%%%%%%%%%%%%%%%%%%%%%%%%%%%%%%%%%%%%%%%%%%%%%%%%%%%%%%%%%%%%%%%%%%%%%%%%%%
\begin{figure*}[hbt]
\normalsize
\hrulefill
\begin{eqnarray}
%%%%%%%%%%%%%%%%%%%%%%%%%%%%%%%%%%%%%%%%%%%%%%%%%%%%%%%%%%%%%%%%%%%%%%%%%%%%%%
%------------------------------\rho 1-- eta_1----------------------------------
\Pi_{\eta_1\eta_1} &=& \int^{s_0}_{16 m_s^2} \Bigg [
{s^4 \over 18432 \pi^6}
- { m_s^2 s^3 \over 256 \pi^6 }
+ \Big ( - { \langle g^2 G G \rangle \over 18432 \pi^6 }
+ {m_s \langle \bar s s \rangle \over 48 \pi^4} \Big ) s^2
\\ \nonumber &&
+ \Big ( { \langle \bar s s \rangle^2 \over 18 \pi^2 }
- { m_s \langle g \bar s \sigma G s \rangle \over 32 \pi^4 }
+ { 17 m_s^2 \langle g^2 G G \rangle \over 9216 \pi^6 } \Big ) s
+ \Big ( { \langle \bar s s \rangle \langle g \bar s \sigma G s \rangle \over 8 \pi^2 }
- { m_s \langle g^2 G G \rangle \langle \bar s s \rangle \over 128 \pi^4}
- { 29 m_s^2 \langle \bar s s \rangle^2 \over 12 \pi^2 } \Big ) \Bigg ] e^{-s/M_B^2} ds
\\ \nonumber &&
+ \Big (  {5 \langle g^2 GG \rangle \langle \bar s s \rangle^2 \over 864 \pi^2}
+ { \langle g \bar s \sigma G s \rangle^2 \over 24 \pi^2 }
+ {20 m_s \langle \bar s s \rangle^3 \over 9}
- {5 m_s \langle g^2 GG \rangle \langle g \bar s \sigma G s \rangle \over 2304 \pi^4 }
- { 13 m_s^2 \langle \bar s s \rangle \langle g \bar s \sigma G s \rangle \over 8 \pi^2 } \Big )
\\ \nonumber &&
+ {1 \over M_B^2} \Big ( - { 32 g^2 \langle \bar s s \rangle^4 \over 81 }
- { \langle g^2 GG \rangle \langle \bar s s \rangle \langle g \bar s \sigma G s \rangle \over 576 \pi^2 }
- { 19 m_s \langle \bar s s \rangle^2 \langle g \bar s \sigma G s \rangle \over 18 }
+ { m_s^2 \langle g^2 GG \rangle \langle \bar s s \rangle^2 \over 576 \pi^2 }
+ { m_s^2 \langle g \bar s \sigma G s \rangle^2 \over 16 \pi^2 } \Big )
\label{eq:pieta1} \, ,
%------------------------------\rho 2-- eta_2----------------------------------
\\ \Pi_{\eta_2\eta_2} &=& \int^{s_0}_{16 m_s^2} \Bigg [
{s^4 \over 12288 \pi^6}
- { 3 m_s^2 s^3 \over 512 \pi^6 }
+ \Big ( { \langle g^2 G G \rangle \over 18432 \pi^6 }
+ {m_s \langle \bar s s \rangle \over 32 \pi^4} \Big ) s^2
\\ \nonumber &&
+ \Big ( { \langle \bar s s \rangle^2 \over 12 \pi^2 }
- { m_s \langle g \bar s \sigma G s \rangle \over 24 \pi^4 }
+ { 35 m_s^2 \langle g^2 G G \rangle \over 9216 \pi^6 } \Big ) s
+ \Big ( { \langle \bar s s \rangle \langle g \bar s \sigma G s \rangle \over 6 \pi^2 }
- { 3 m_s \langle g^2 G G \rangle \langle \bar s s \rangle \over 128 \pi^4}
- { 29 m_s^2 \langle \bar s s \rangle^2 \over 8 \pi^2 }  \Big ) \Bigg ] e^{-s/M_B^2} ds
\\ \nonumber &&
+ \Big ( {5 \langle g^2 GG \rangle \langle \bar s s \rangle^2 \over 288 \pi^2}
+ { 5 \langle g \bar s \sigma G s \rangle^2 \over 96 \pi^2 }
+ {10 m_s \langle \bar s s \rangle^3 \over 3}
- {5 m_s \langle g^2 GG \rangle \langle g \bar s \sigma G s \rangle \over 768 \pi^4 }
- { 19 m_s^2 \langle \bar s s \rangle \langle g \bar s \sigma G s \rangle \over 8 \pi^2 } \Big )
\\ \nonumber &&
+ {1 \over M_B^2} \Big ( - { 16 g^2 \langle \bar s s \rangle^4 \over 27 }
- { \langle g^2 GG \rangle \langle \bar s s \rangle \langle g \bar s \sigma G s \rangle \over 192 \pi^2 }
- { 29 m_s \langle \bar s s \rangle^2 \langle g \bar s \sigma G s \rangle \over 18 }
- { m_s^2 \langle g^2 GG \rangle \langle \bar s s \rangle^2 \over 576 \pi^2 }
+ { 5 m_s^2 \langle g \bar s \sigma G s \rangle^2 \over 48 \pi^2 } \Big )
\label{eq:pieta2} \, ,
\\ \Pi_{\eta_1\eta_2} &=& i \int^{s_0}_{16 m_s^2} \Bigg [
{ \langle g^2 G G \rangle \over 6144 \pi^6 } s^2
+ { 3 m_s^2 \langle g^2 G G \rangle \over 1024 \pi^6 } s
- { 3 m_s \langle g^2 G G \rangle \langle \bar s s \rangle \over 128 \pi^4} \Bigg ] e^{-s/M_B^2} ds
\\ \nonumber &&
+ i \Big ( {5 \langle g^2 GG \rangle \langle \bar s s \rangle^2 \over 288 \pi^2}
- {5 m_s \langle g^2 GG \rangle \langle g \bar s \sigma G s \rangle \over 768 \pi^4 } \Big )
+ {i \over M_B^2} \Big ( - { \langle g^2 GG \rangle \langle \bar s s \rangle \langle g \bar s \sigma G s \rangle \over 192 \pi^2 }
- { m_s^2 \langle g^2 GG \rangle \langle \bar s s \rangle^2 \over 192 \pi^2 } \Big )
\label{eq:eta12} \, .
\end{eqnarray}
\hrulefill
\vspace*{4pt}
\end{figure*}
%%%%%%%%%%%%%%%%%%%%%%%%%%%%%%%%%%%%%%%%%%%%%%%%%%%%%%%%%%%%%%%%%%%%%%%%%%%%%%

\begin{figure*}[]
\begin{center}
\includegraphics[width=0.4\textwidth]{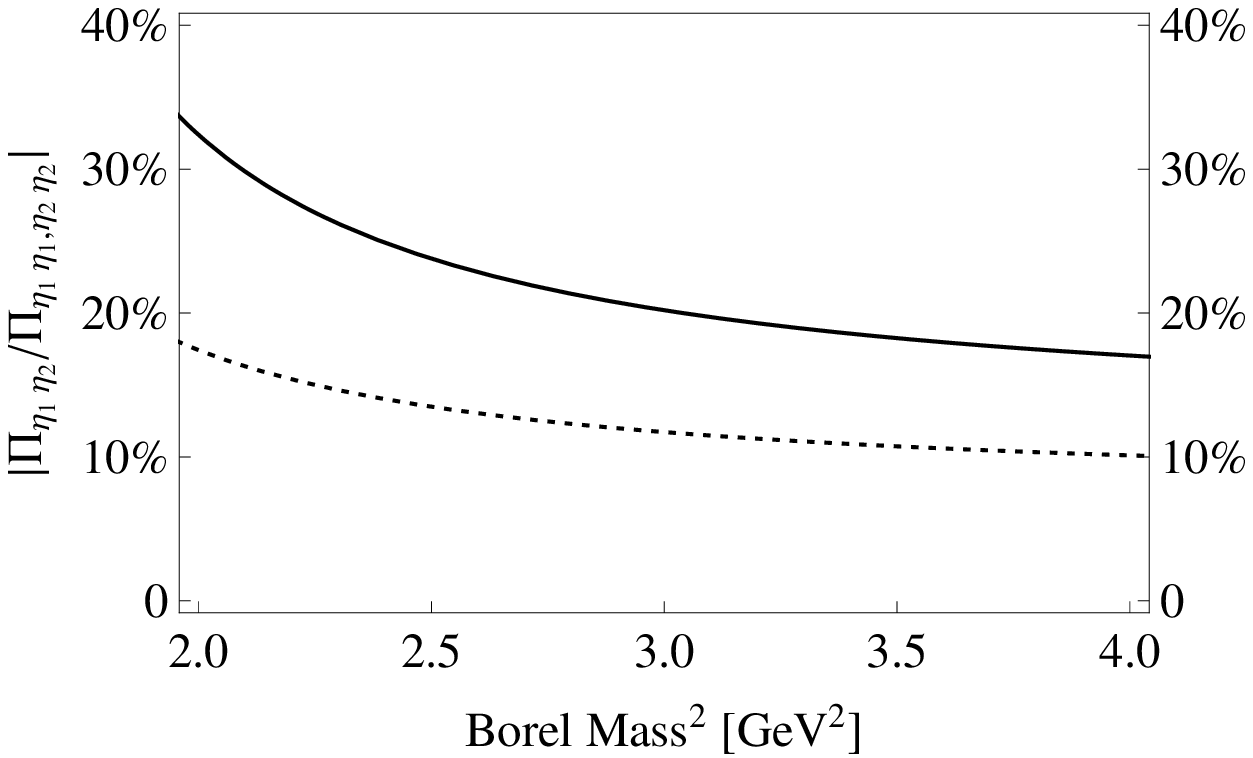}
\includegraphics[width=0.4\textwidth]{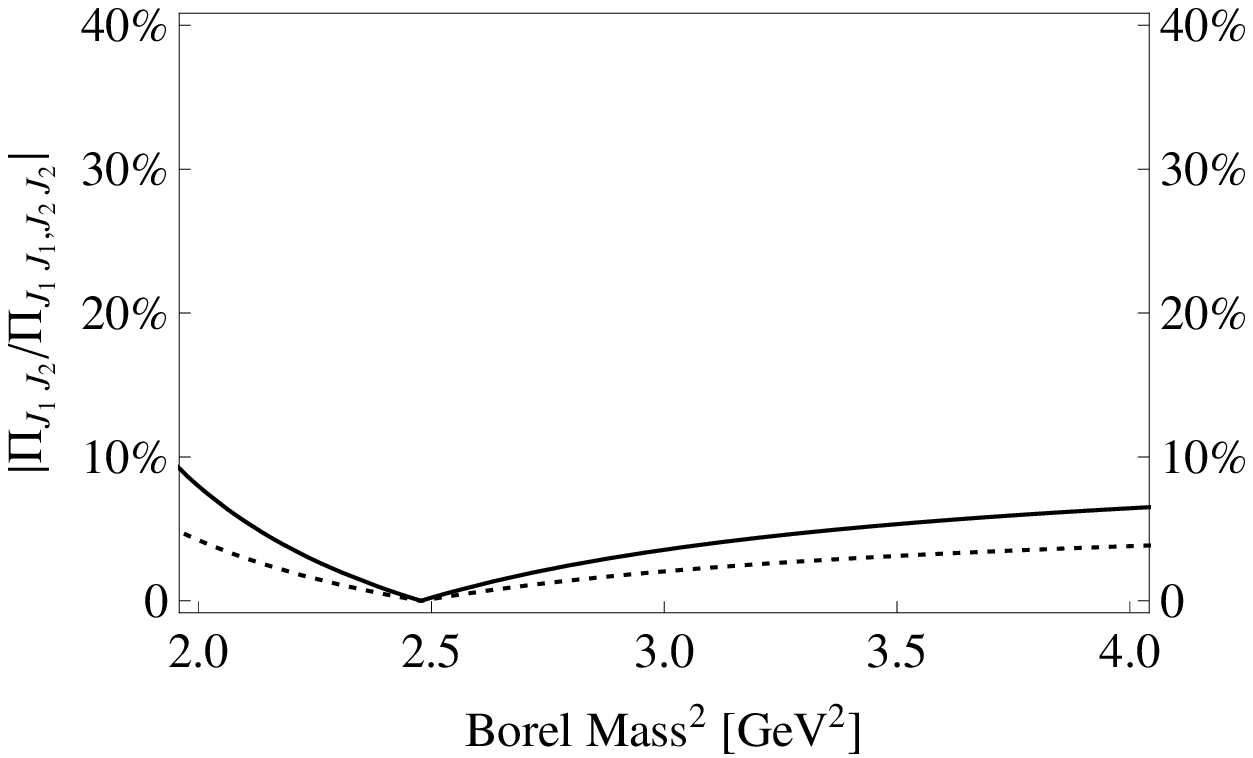}
\end{center}
\caption{Left: $\left|{\Pi_{\eta_1\eta_2}(M_B^2)\over\Pi_{\eta_1\eta_1}(M_B^2)}\right|$ (solid) and $\left|{\Pi_{\eta_1\eta_2}(M_B^2)\over\Pi_{\eta_2\eta_2}(M_B^2)}\right|$ (dotted), as functions of the Borel mass $M_B$, when taking $s_0 = 6.0$ GeV$^2$.
Right: $\left|{\Pi_{J_1J_2}(M_B^2)\over\Pi_{J_1J_1}(M_B^2)}\right|$ (solid) and $\left|{\Pi_{J_1J_2}(M_B^2)\over\Pi_{J_2J_2}(M_B^2)}\right|$ (dotted), as functions of the Borel mass $M_B$, when taking $s_0 = 6.0$ GeV$^2$.
}
\label{fig:offdiagonal}
\end{figure*}

After fixing $s_0 = 6.0$ GeV$^2$, we show $\Pi_{\eta_1\eta_2}(M_B^2)$ as a function of the Borel mass $M_B$ in the left panel of Fig.~\ref{fig:offdiagonal}, compared with $\Pi_{\eta_1\eta_1}(M_B^2)$ and $\Pi_{\eta_2\eta_2}(M_B^2)$. Especially, we have
\begin{equation}
\left|{\Pi_{\eta_1\eta_2}(3.0~{\rm GeV}^2) \over \Pi_{\eta_1\eta_1}(3.0~{\rm GeV}^2)}\right| = 0.20 \, , \, \left|{\Pi_{\eta_1\eta_2}(3.0~{\rm GeV}^2) \over \Pi_{\eta_2\eta_2}(3.0~{\rm GeV}^2)}\right| = 0.12 \, .
\end{equation}
These values suggest that the off-diagonal term is non-ignorable. By diagonalizing the following matrix at around $s_0= 6.0$ GeV$^2$ and $M_B^2= 2.5$ GeV$^2$
\begin{eqnarray}
\left( \begin{array}{cc}
\Pi_{\eta_1\eta_1} & \Pi_{\eta_1\eta_2}
\\ \Pi_{\eta_1\eta_2}^\dagger & \Pi_{\eta_2\eta_2}
\end{array} \right) \, ,
\end{eqnarray}
we obtain two new currents $J_{1\mu}$ and $J_{2\mu}$ with the mixing angle $\theta = -5.0^{\rm o}$, which do not strongly correlate to each other. Again we fix $s_0 = 6.0$ GeV$^2$, and show $\Pi_{J_1J_2}(M_B^2)$ as a function of the Borel mass $M_B$ in the right panel of Fig.~\ref{fig:offdiagonal}, compared with $\Pi_{J_1J_1}(M_B^2)$ and $\Pi_{J_2J_2}(M_B^2)$. Especially, we have
\begin{equation}
\left|{\Pi_{J_1J_2}(3.0~{\rm GeV}^2) \over \Pi_{J_1J_1}(3.0~{\rm GeV}^2)}\right| = 0.04 \, , \,  \left|{\Pi_{J_1J_2}(3.0~{\rm GeV}^2) \over \Pi_{J_2J_2}(3.0~{\rm GeV}^2)}\right| = 0.02 \, .
\end{equation}

%
%=====================================================================================
%=====================================================================================
\section{Numerical Analysis}
\label{sec:numerical}
%=====================================================================================
%=====================================================================================
%

\begin{figure}[hbt]
\begin{center}
\includegraphics[width=0.4\textwidth]{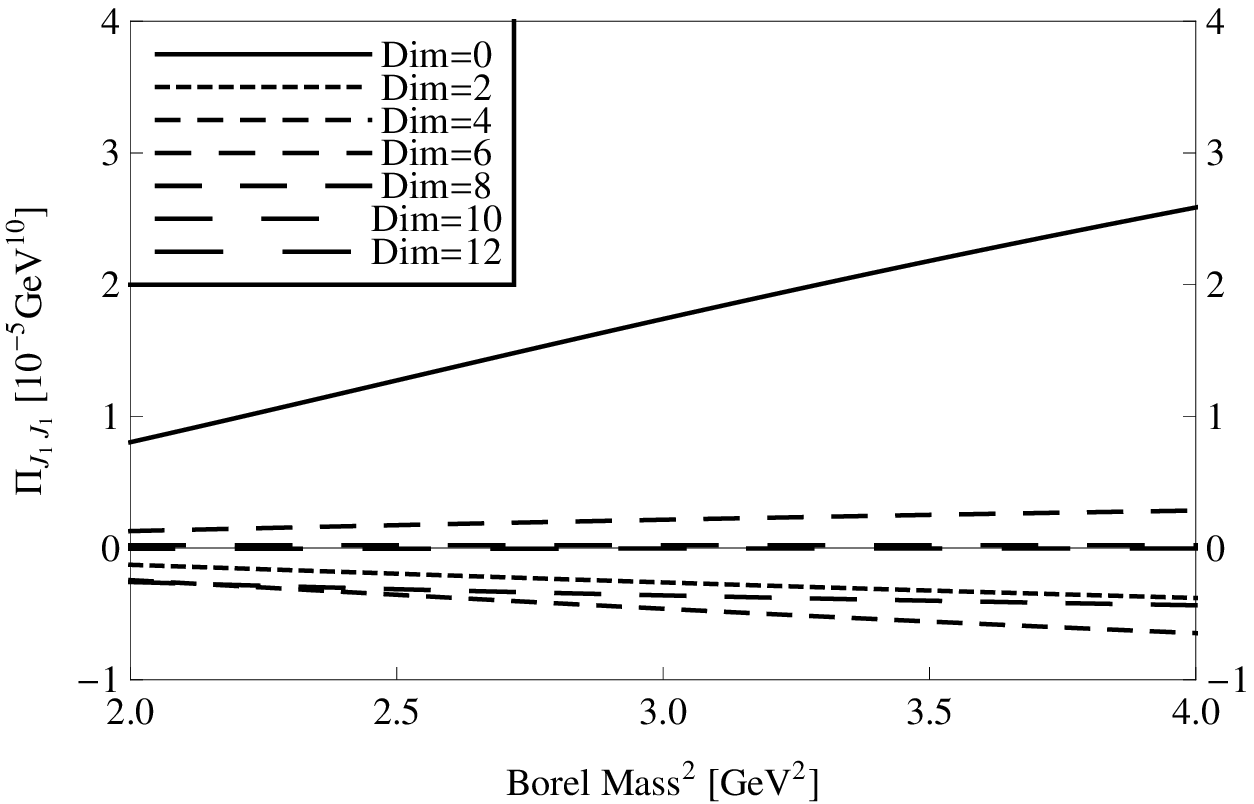}
\caption{Various contributions to the correlation function $\Pi_{J_1J_1}(M_B^2)$, as functions of the Borel mass $M_B$ in unit of GeV$^{10}$, when taking $s_0 = 6.0$ GeV$^2$.}
\label{fig:convergence}
\end{center}
\end{figure}

In this section we use the currents $J_{1\mu}$ and $J_{2\mu}$ to perform QCD sum rule analyses. Take $J_{1\mu}$ as an example. First we study the convergence of the operator product expansion, which is the cornerstone of the reliable QCD sum rule analysis. %We fix the integral subscript $16 m_s^2$ to be zero and superscript $s_0$ to be $\infty$, and obtain the numerical series of the OPE as functions of $M_B$:
%\begin{eqnarray}
%\nonumber \Pi_1(M_B^2) &=& 1.4 \times 10^{-6} M_B^{10} - 2.3 \times 10^{-7} M_B^8
%\\ \nonumber && - 5.2 \times 10^{-7} M_B^6 + 4.0 \times 10^{-7} M_B^4
%\\ \nonumber && - 1.5 \times 10^{-6} M_B^2 + 2.1 \times 10^{-7}
%\\ && - 1.2 \times 10^{-7} M_B^{-2} \, .
%\\ \nonumber
%\Pi_2(M_B^2) &=& 2.0 \times 10^{-6} M_B^{10} - 3.4 \times 10^{-7} M_B^8
%\\ \nonumber && - 6.4 \times 10^{-7} M_B^6 + 6.5 \times 10^{-7} M_B^4
%\\ \nonumber && - 1.9 \times 10^{-6} M_B^2 + 2.6 \times 10^{-7}
%\\ && - 1.7 \times 10^{-7} M_B^{-2} \, .
%\end{eqnarray}
%From this equation, we find that the convergence of the OPE is satisfied when $M_B^2$ is larger than 2 GeV$^2$. Based on this,
To do this we require that the $D=10$ and $D=12$ terms be less than 5\%:
\begin{eqnarray}
\mbox{CVG} &\equiv& \left|\frac{ \Pi^{D=8+10}(M_B^2) }{ \Pi(M_B^2) }\right| \leq 5\% \, .
\label{eq_convergence}
\end{eqnarray}
After fixing $s_0 = 6.0$ GeV$^2$, we find that this condition is satisfied when $M_B^2$ is larger than 2.0 GeV$^2$. We also show the relative contribution of each term to the correlation function $\Pi_{J_1J_1}(M_B^2)$ in Fig.~\ref{fig:convergence}. We find that in the region 2.0 GeV$^2 < M_B^2 < 4.0$ GeV$^2$, the perturbative term ($D=0$) gives the most important contribution, and the convergence is quite good.
%
%%%%%%%%%%%%%%%%%%%%%%%%%%%%%%%%%%%%%%%%%%%%%%%%%%%%%%%%%%%%%%%%%%%%%%%%%%%%%%
%---------figure current 1
%%%%%%%%%%%%%%%%%%%%%%%%%%%%%%%%%%%%%%%%%%%%%%%%%%%%%%%%%%%%%%%%%%%%%%%%%%%%%%
%

A common problem, when studying multiquark states using QCD sum rules, is how to differentiate the multiquark state and the relevant threshold, because the interpolating current can couple to both of them. For the case of the $Y(2175)$, its relevant threshold is the $\phi f_0(980)$ around 2.0 GeV, which $J_{1\mu}$ and $J_{2\mu}$ can both couple to. Moreover, the $Y(2175)$ is not the lowest state in the $1^{--}$ channel containing $s\bar s$, and $J_{1\mu}$ and $J_{2\mu}$ may also couple to the $\phi(1680)$ (for example, see the Belle experiment~\cite{Shen:2009zze} observing the $\phi(1680)$ and $Y(2175)$ at the same time).

%
%%%%%%%%%%%%%%%%%%%%%%%%%%%%%%%%%%%%%%%%%%%%%%%%%%%%%%%%%%%%%%%%%%%%%%%%%%%%%%
%---------figure current 1
\begin{figure}[hbt]
\begin{center}
\includegraphics[width=0.4\textwidth]{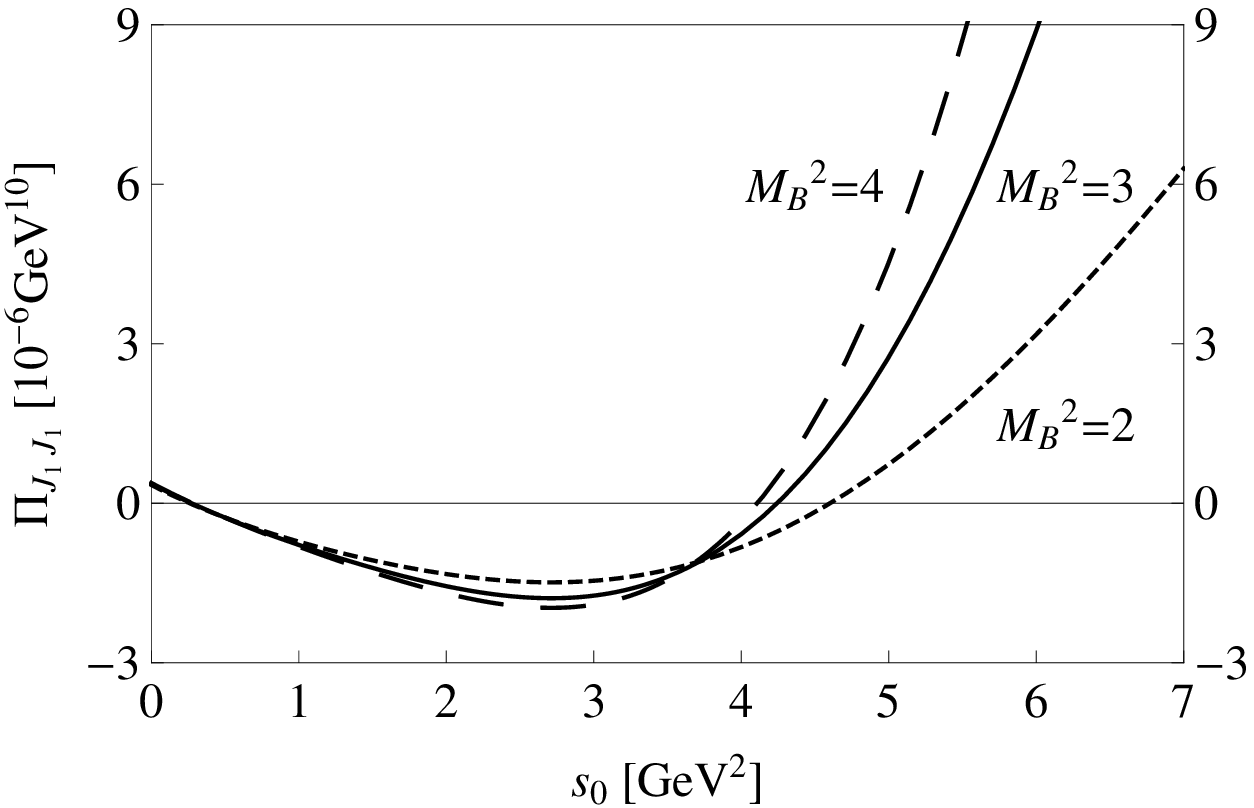}
\caption{The correlation function $\Pi_{J_1J_1}(M_B^2)$ as a function of $s_0$ in unit of GeV$^{10}$. The curves are obtained by taking $M_B^2 = 2.0$ GeV$^2$ (short-dashed), 3.0 GeV$^2$ (solid), and 4.0 GeV$^2$ (long-dashed).}
\label{fig:pi1}
\end{center}
\end{figure}
%%%%%%%%%%%%%%%%%%%%%%%%%%%%%%%%%%%%%%%%%%%%%%%%%%%%%%%%%%%%%%%%%%%%%%%%%%%%%%
%

If this happens, the resulting correlation function should be positive.
Fortunately, we find that the correlation functions $\Pi_{J_1J_1}(M_B^2)$ and $\Pi_{J_2J_2}(M_B^2)$ are negative, and so non-physical, in the region $s_0 < 4.0$ GeV$^2$ when taking $2.0$ GeV$^2 < M_B^2 < 4.0$ GeV$^2$. As an illustration, we show the correlation function $\Pi_{J_1J_1}(M_B^2)$ as a function of $s_0$ in Fig.~\ref{fig:pi1} for $M_B^2 = 2.0/3.0/4.0$ GeV$^2$. This fact indicates that $J_{1\mu}$ and $J_{2\mu}$ both couple weakly to the lower state $\phi(1680)$ as well as the $\phi f_0(980)$ threshold, so the states they couple to, as if they can couple to some states, should be new and possibly exotic states. However, due to the above negative contributions to the correlation functions, the pole contribution is not large enough. This small pole contribution also suggests that the continuum contribution is important, which demands a careful choice of the parameters of QCD sum rules. Accordingly, in the present study we require that the extracted mass have a dual minimum dependence on both the threshold value $s_0$ and the Borel mass $M_B$.

%
%%%%%%%%%%%%%%%%%%%%%%%%%%%%%%%%%%%%%%%%%%%%%%%%%%%%%%%%%%%%%%%%%%%%%%%%%%%%%%
%---------figure current 1
\begin{figure*}[]
\begin{center}
\includegraphics[width=0.4\textwidth]{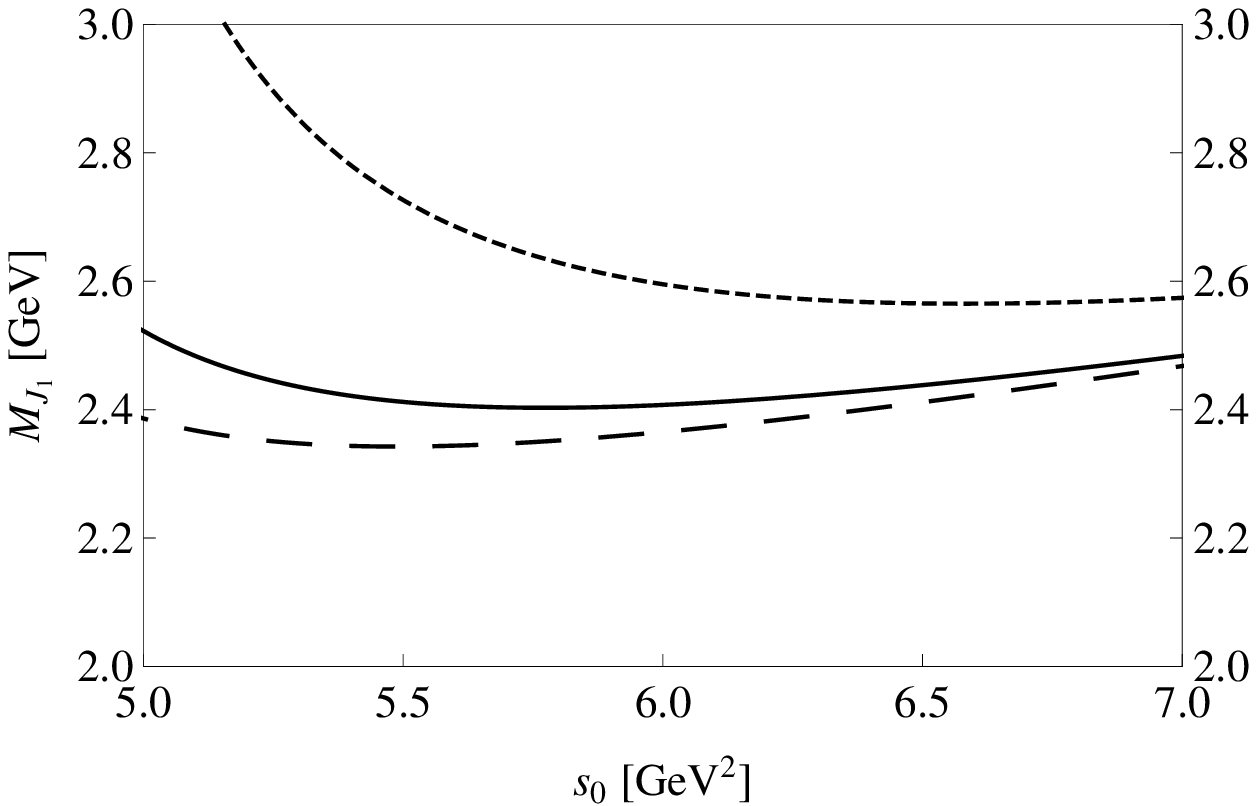}
\includegraphics[width=0.4\textwidth]{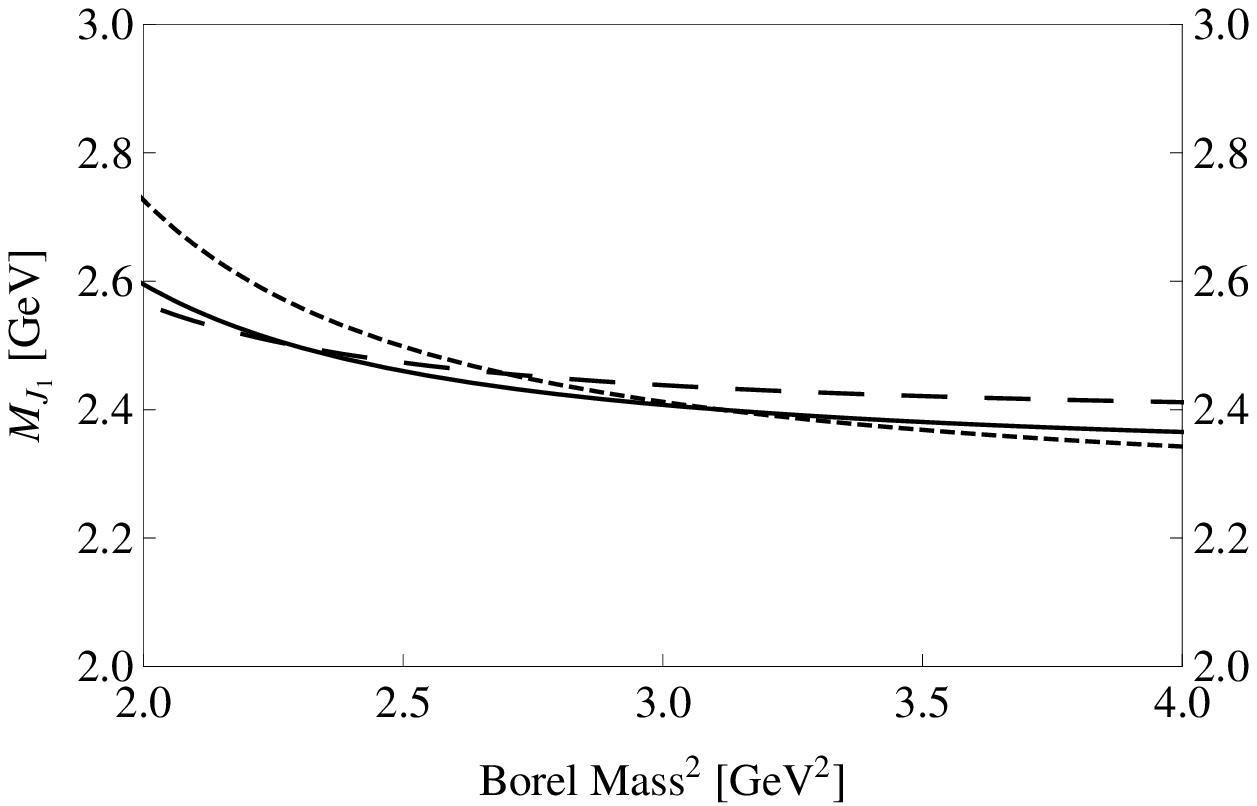} \caption{
Mass calculated using the current $J_{1\mu}$, as a function of the threshold value $s_0$ (left) and the Borel mass $M_B$ (right).
In the left panel, the short-dashed/solid/long-dashed curves are obtained by setting $M_B^2 = 2.0/3.0/4.0$ GeV$^2$, respectively.
In the right panel, the short-dashed/solid/long-dashed curves are obtained by setting $s_0 = 5.5/6.0/6.5$ GeV$^2$, respectively.}
\label{fig:J1mass}
\end{center}
\end{figure*}
%%%%%%%%%%%%%%%%%%%%%%%%%%%%%%%%%%%%%%%%%%%%%%%%%%%%%%%%%%%%%%%%%%%%%%%%%%%%%%
%
Still using $J_{1\mu}$ as an example, we show the mass obtained using Eq.~(\ref{eq_LSR}) as a function of the threshold value $s_0$ and the Borel mass $M_B$ in Fig.~\ref{fig:J1mass}. We find that there is a mass minimum at around 2.4 GeV when taking $s_0$ to be around $6.0$ GeV$^2$, and at the same time the Borel mass dependence is weak at around 3.0 GeV$^2$. Accordingly, we fix $s_0$ to be around $6.0$ GeV$^2$ and $M_B^2$ to be around 3.0 GeV$^2$, and choose our working regions to be 5.0 GeV$^2< s_0 < 7.0$ GeV$^2$ and 2.0 GeV$^2 < M_B^2 < 4.0$ GeV$^2$. These regions are moderately large enough for the mass prediction, where the mass is extracted to be
\begin{equation}
M_{Y_1} = 2.41 \pm 0.25 {\rm~GeV} \, .
\end{equation}
Here the uncertainty is due to the Borel mass $M_B$, the threshold value $s_0$, and various condensates~\cite{Yang:1993bp,Narison:2002pw,Gimenez:2005nt,Jamin:2002ev,Ioffe:2002be,Ovchinnikov:1988gk,Ellis:1996xc,pdg}.

%
%%%%%%%%%%%%%%%%%%%%%%%%%%%%%%%%%%%%%%%%%%%%%%%%%%%%%%%%%%%%%%%%%%%%%%%%%%%%%%
%---------figure current 2
\begin{figure*}[]
\begin{center}
\includegraphics[width=0.4\textwidth]{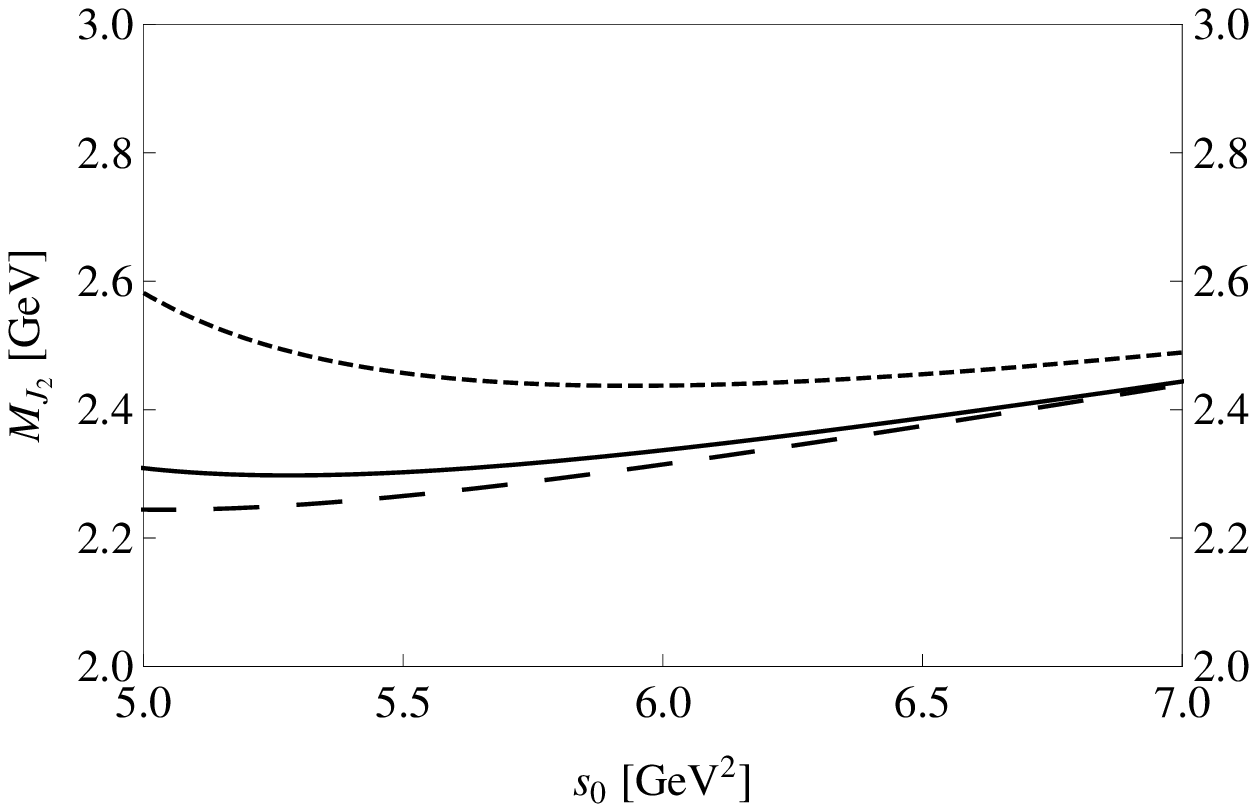}
\includegraphics[width=0.4\textwidth]{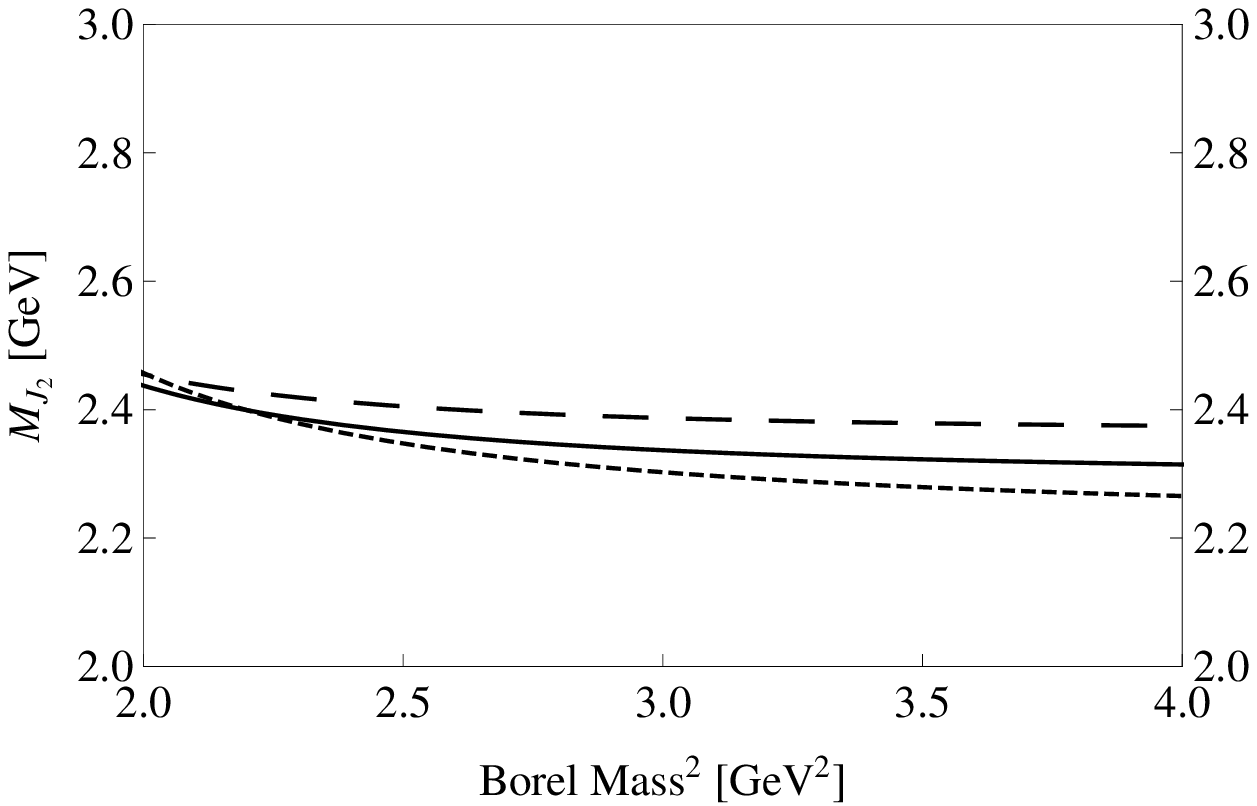} \caption{
Mass calculated using the current $J_{2\mu}$, as a function of the threshold value $s_0$ (left) and the Borel mass $M_B$ (right).
In the left panel, the short-dashed/solid/long-dashed curves are obtained by setting $M_B^2 = 2.0/3.0/4.0$ GeV$^2$, respectively.
In the right panel, the short-dashed/solid/long-dashed curves are obtained by setting $s_0 = 5.5/6.0/6.5$ GeV$^2$, respectively.}
\label{fig:J2mass}
\end{center}
\end{figure*}
%%%%%%%%%%%%%%%%%%%%%%%%%%%%%%%%%%%%%%%%%%%%%%%%%%%%%%%%%%%%%%%%%%%%%%%%%%%%%%
%

Similarly, we use $J_{2\mu}$ to perform QCD sum rule analyses. Choosing the same working regions 5.0 GeV$^2< s_0 < 7.0$ GeV$^2$ and 2.0 GeV$^2 < M_B^2 < 4.0$ GeV$^2$, the mass is extracted to be
\begin{equation}
M_{Y_2} = 2.34 \pm 0.17 {\rm~GeV} \, .
\end{equation}
The above result is shown in Fig.~\ref{fig:J2mass} as a function of the threshold value $s_0$ and the Borel mass $M_B$.

%
%%%%%%%%%%%%%%%%%%%%%%%%%%%%%%%%%%%%%%%%%%%%%%%%%%%%%%%%%%%%%%%%%%%%%%%%%%%%%%
%---------figure current 1
\begin{figure*}[]
\begin{center}
\includegraphics[width=0.4\textwidth]{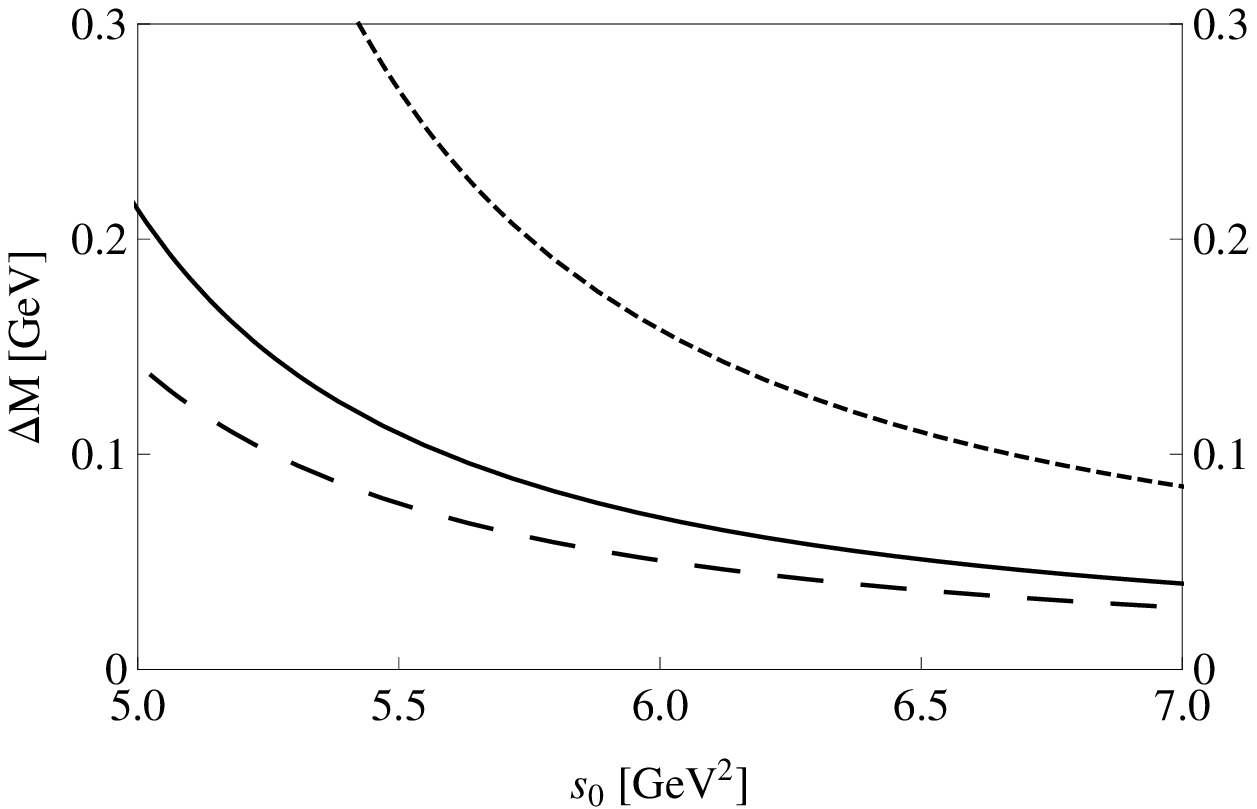}
\includegraphics[width=0.4\textwidth]{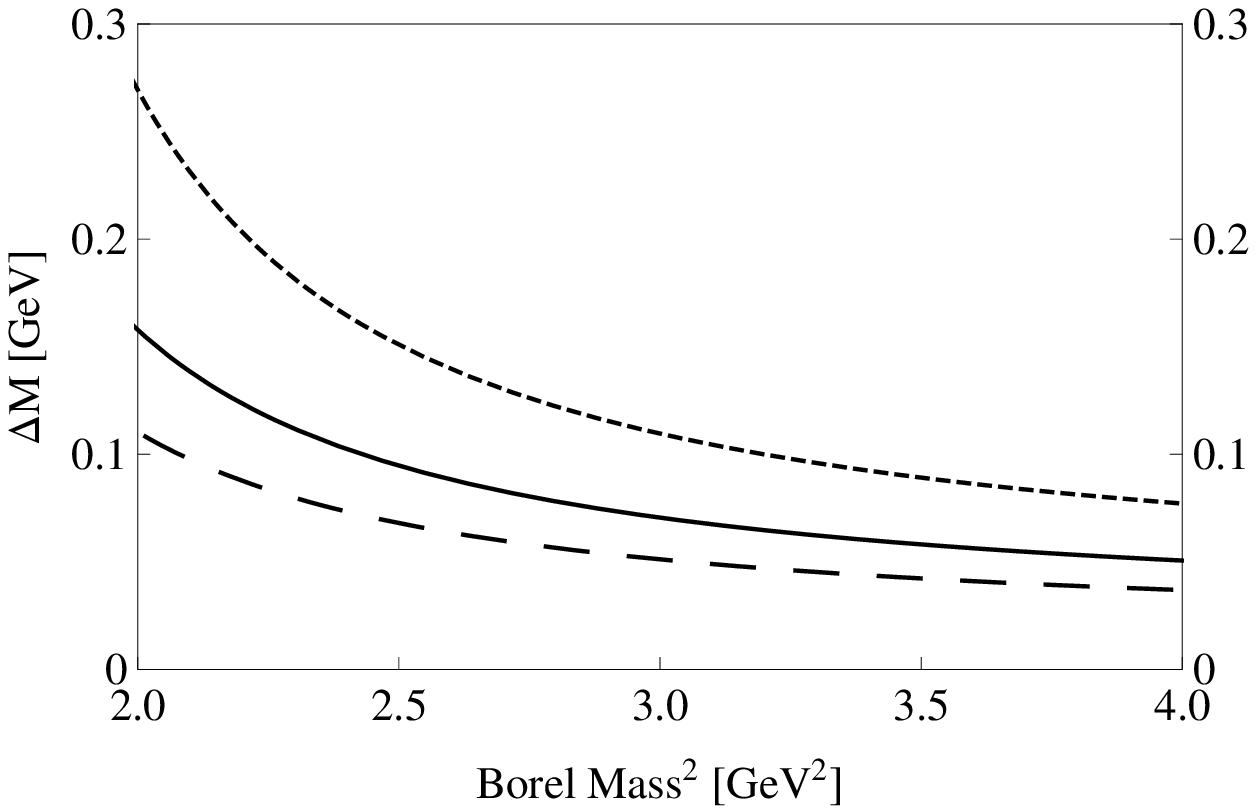} \caption{
Mass splitting between the two currents $J_{1\mu}$ and $J_{2\mu}$, as a function of the threshold value $s_0$ (left) and the Borel mass $M_B$ (right).
In the left panel, the short-dashed/solid/long-dashed curves are obtained by setting $M_B^2 = 2.0/3.0/4.0$ GeV$^2$, respectively.
In the right panel, the short-dashed/solid/long-dashed curves are obtained by setting $s_0 = 5.5/6.0/6.5$ GeV$^2$, respectively.}
\label{fig:splitting}
\end{center}
\end{figure*}
%%%%%%%%%%%%%%%%%%%%%%%%%%%%%%%%%%%%%%%%%%%%%%%%%%%%%%%%%%%%%%%%%%%%%%%%%%%%%%
%

As we have discussed in previous sections, $J_{1\mu}$ and $J_{2\mu}$ may couple to two different physical states. Using the same working region, we evaluate the mass splitting between these two states/currents to be
\begin{equation}
\Delta M = 71 ^{+172}_{-~48}  {\rm~MeV} \, .
\end{equation}
The above result is shown in Fig.~\ref{fig:splitting} as a function of the threshold value $s_0$ and the Borel mass $M_B$.

%
%=====================================================================================
%=====================================================================================
\section{Summary and Discussions}
\label{sec:summary}
%=====================================================================================
%=====================================================================================
%

In this work we apply the method of QCD sum rules to study the $Y(2175)$ by using local $ss\bar s\bar s$ interpolating currents with $J^{PC} = 1^{--}$. The relevant diquark-antidiquark $(ss)(\bar s \bar s)$ and meson-meson $(\bar s s)(\bar s s)$ interpolating currents have been systematically constructed in Ref.~\cite{Chen:2008ej}, where their relations have also been derived. There we found two independent currents, so there are (at least) two different internal structures.
In Ref.~\cite{Chen:2008ej} we have calculated the two diagonal terms using the two diquark-antidiquark $(ss)(\bar s \bar s)$ currents $\eta_{1\mu}$ and $\eta_{2\mu}$, and in this work we further calculate their off-diagonal term
\begin{eqnarray}
\langle 0 | T \eta_{1\mu}(x) { \eta_{2\nu}^\dagger } (0) | 0 \rangle \, .
\end{eqnarray}
We find two new currents $J_{1\mu}$ and $J_{2\mu}$ with the mixing angle $\theta = -5.0^{\rm o}$:
\begin{eqnarray}
J_{1\mu} &=& \cos\theta~\eta_{1\mu} + \sin\theta~i~\eta_{2\mu} \, ,
\\ \nonumber J_{2\mu} &=& \sin\theta~\eta_{1\mu} + \cos\theta~i~\eta_{2\mu} \, .
\end{eqnarray}
These two currents do not strongly correlate to each other, suggesting that they may couple to different physical states.

We use $J_{1\mu}$ and $J_{2\mu}$ to perform QCD sum rule analyses. Especially, we find that $J_{1\mu}$ and $J_{2\mu}$ both couple weakly to the lower state $\phi(1680)$ as well as the $\phi f_0(980)$ threshold, so the states they couple to, as if they can couple to some states, should be new and possibly exotic states. Accordingly, we assume $J_{1\mu}$ and $J_{2\mu}$ separately couple to two different states with the same quantum number $J^{PC} = 1^{--}$, whose masses are extracted to be
\begin{eqnarray}
M_{Y_1} &=& 2.41 \pm 0.25  {\rm~GeV} \, ,
\\ M_{Y_2} &=& 2.34 \pm 0.17  {\rm~GeV} \, .
\end{eqnarray}
These results do not change significantly compared with those obtained in Ref.~\cite{Chen:2008ej}. However, their mass splitting depend significantly on the mixing angle, and we use $J_{1\mu}$ and $J_{2\mu}$ with $\theta = -5.0^{\rm o}$ to evaluate it to be
\begin{equation}
\Delta M = 71 ^{+172}_{-~48}  {\rm~MeV} \, .
\end{equation}
The mass extracted using $J_{2\mu}$ is consistent with the experimental mass of the $Y(2175)$, suggesting that $J_{2\mu}$ may couple to the $Y(2175)$; while the mass extracted using $J_{1\mu}$ is a bit larger, suggesting that the $Y(2175)$ may have a partner state whose mass is around $71 ^{+172}_{-~48}$ MeV larger.

Because $J_{1\mu}$ and $J_{2\mu}$ are two $ss\bar s\bar s$ interpolating currents with $J^{PC} = 1^{--}$, both the $Y(2175)$ and its possible partner state should be vector mesons containing large strangeness components. Note that our results do not definitely suggest that they are $s s \bar s \bar s$ tetraquark states, because the interpolating current sees only the quantum numbers of the physical state, that is $J^{PC} = 1^{--}$.
We can further use Eq.~(\ref{eq:transform}), which is derived from the Fierz transformation, to obtain that the $Y(2175)$ and its possible partner state can both be observed in the $\phi f_0(980)$ channel, while the latter may also be observed in the $\phi f_1(1420)$ channel, as if kinematically allowed.

Experimentally, the $Y(2175)$ has been well established by the BaBar, BESII, BESIII, and Belle experiments. Besides it, there might be another structure in the $\phi f_0(980)$ invariant mass spectrum at around 2.4 GeV. This might be the partner state of the $Y(2175)$, which is coupled by the current $J_{1\mu}$.
To end this paper, we note that the two mass values we obtained, $2.34 \pm 0.17$ GeV and $2.41 \pm 0.25$ GeV, are both around 2.4 GeV, indicating that there might be even more complicated structures in this region, such as two coherent resonances. We also note that there are many charmonium-like $Y$ states of $J^{PC} = 1^{--}$, so it is natural to think that there can be more than one $Y$ states in the light sector.
Accordingly, we propose to carefully study the structure in the $\phi f_0(980)$ invariant mass spectrum at around 2.4 GeV in future experiments.

%
%=====================================================================================
%=====================================================================================
%=====================================================================================
\section*{Acknowledgments}
%=====================================================================================
%=====================================================================================
%=====================================================================================
%

This project is supported by
the National Natural Science Foundation of China under Grants No. 11475015, No. 11575008, No. 11575017, No. 11722540, No. 11261130311, and No. 11761141009,
the National Key Basic Research Program of China (2015CB856700),
the Fundamental Research Funds for the Central Universities,
and the Foundation for Young Talents in College of Anhui Province (Grants No. gxyq2018103)..

%
%%%%%%%%%%%%%%%%%%%%%%%%%%%%%%%%%%%%%%%%%%%%%%%%%%%%%%%%%%%%%%%%%%%%%%%%%%%%%%

%%%%%%%%%%%%%%%%%%%%%%%%%%%%%%%%%%%%%%%%%%%%%%%%%%%%%%%%%%%%%%%%%%%%%%%%%%%%%%
%

\end{document}